\documentclass[11pt]{article}
\usepackage[USenglish]{babel}
\usepackage{a4wide}
\usepackage{epsfig}
\usepackage[all]{xy}
\usepackage{amssymb}
\usepackage{amsmath}
\usepackage{latexsym}
\usepackage{euler}
\usepackage{pifont}
\usepackage{pdfsync}
\usepackage{enumerate}
\usepackage[applemac]{inputenc} 
\usepackage[OT1]{fontenc}
\title{A Stochastic Interpretation of  Game Logic}
\author{Ernst-Erich Doberkat\thanks{Funded in part by Deutsche Forschungsgemeinschaft, grant DO 263/12-1, \emph{Koalgebraische Eigenschaften stochastischer Relationen.}}\\
Chair for Software Technology\\
Technische Universitt Dortmund\\
\texttt{ernst-erich.doberkat@udo.edu}}
\date{\today}
\date{\today}
\newcommand{\labelImpl}[2]{\ensuremath{\ref{#1}~\Rightarrow~\ref{#2}}}

\newcommand{\Inv}[2]{\ensuremath{{\mathcal INV}\left(#1, #2\right)}}
\newcommand{\Klasse}[2]{\left[#1\right]_{#2}}
\newcommand{\Faktor}[2]{{#1}/{#2}}
\newcommand{\fMap}[1]{\eta_{#1}}
\newcommand{\Bild}[2]{{#1}\left[#2\right]}
\newcommand{\InvBild}[2]{\Bild{#1^{-1}}{#2}}
\newcommand{\Kern}[1]{\mathsf{ker}\left(#1\right)}
\newcommand{\Folge}[1]{(#1_n)_{n \in \Nat}}

\newcommand{\theTheory}[2]{Th_{#1}({#2})}
\newcommand{\spaceFont}[1]{\mathfrak{#1}}

\newcommand{\SubProb}[1]{\spaceFont{S}\left(#1\right)}

\newcommand{\SubProbSenza}{\spaceFont{S}}

\newcommand{\PowerSet}[1]{\ensuremath{\mathcal{P}\left(#1\right)}}

\newcommand{\Borel}[1]{\ensuremath{{\mathcal B}(#1)}}

\edef\LinkeKlammer{\lbrack\!\lbrack}
\edef\RechteKlammer{\rbrack\!\rbrack}
\newcommand{\Gilt}[1][\phi]{\ensuremath{\LinkeKlammer#1\RechteKlammer}}

\newcommand{\Trans}{\rightsquigarrow}

\newtheorem{definition}{Definition}[section]
\newcommand{\BeginDefinition}[1]{%
  \begin{definition}\label{#1}
}
\newcommand{\EndDefinition}{\end{definition}}

\newtheorem{example}[definition]{Example}
\newcommand{\BeginExample}[1]{%
  \begin{example}\label{#1}\rm
}
\newcommand{\EndExample}{--- \end{example}}

\newtheorem{observation}[definition]{Observation}
\newcommand{\BeginObservation}[1]{
  \begin{observation}\label{#1}\rm
}
\newcommand{\EndObservation}{--- \end{observation}}

\newtheorem{theorem}[definition]{Theorem}
\newcommand{\BeginTheorem}[1]{%
  \begin{theorem}\label{#1}
}
\newcommand{\EndTheorem}{\end{theorem}}

\newtheorem{corollary}[definition]{Corollary}
\newcommand{\BeginCorollary}[1]{
  \begin{corollary}\label{#1}
}

\newtheorem{proposition}[definition]{Proposition}
\newcommand{\BeginProposition}[1]{%
  \begin{proposition}\label{#1}
}
\newcommand{\EndProposition}{\end{proposition}}

\newcommand{\EndCorollary}{\end{corollary}}
\newtheorem{lemma}[definition]{Lemma}
\newcommand{\BeginLemma}[1]{%
  \begin{lemma}\label{#1}
}
\newcommand{\EndLemma}{\end{lemma}}

\newtheorem{claim}{Claim}
\newcommand{\BeginClaim}[1]{%
  \begin{claim}\label{#1}
}
\newcommand{\EndClaim}{\end{claim}}

\newenvironment{proof}{\textbf{Proof\ }}{\ensuremath{\QED}}
\newcommand{\BeginProof}{\begin{proof}}
\newcommand{\EndProof}{\end{proof}}

\newenvironment{remark}{\textbf{Remark:\ }}{}
\newcommand{\BeginRemark}{\begin{remark}}
\newcommand{\EndRemark}{\QED\end{remark}}
\newcommand{\QED}{%
\ensuremath{\dashv}
}

\newcommand{\Real}{\mathbb{R}}
\newcommand{\pReal}{\mathbb{R}_{+}}
\newcommand{\Nat}{\mathbb{N}}
\newcommand{\Rational}{\mathbb{Q}}

\usepackage{fancyhdr}
\renewcommand{\Borel}[1]{\ensuremath{{\spaceFont B}(#1)}}
\renewcommand{\spaceFont}{\mathcal}
\def\VauSenza{\ensuremath{\spaceFont{V}}}
\newcommand{\Vau}[1]{\ensuremath{\ensuremath{\VauSenza(#1)}}}
\newcommand{\bas}[2]{\ensuremath{\beta_{#1, #2}}}
\renewcommand{\bas}[2]{\ensuremath{{\boldsymbol\beta}(#1, #2)}}
\newcommand{\basS}[3]{\bas{#2}{#3 \ast #1}}
\renewcommand{\basS}[3]{\bas{#2}{#3\,\pmb{\mid}\,#1}}
\renewcommand{\basS}[3]{\ensuremath{{\boldsymbol\beta}_{#1}(#2, #3)}}
\newcommand{\Meas}[1]{\spaceFont{M}\left(#1\right)}
\newcommand{\iReal}{\widetilde{\mathbb{R}}_+}
\def\Ge{\mathcal{G}}
\def\He{\mathcal{H}}
\def\Ke{\mathcal{K}}

\def\Me{\mathcal{M}}
\newcommand{\dasI}[4][\mathcal{G}]{\ensuremath{\Im_{#1}(#2 \mid #3, #4)}}
\newcommand{\dasL}[3]{\ensuremath{\mathcal{L}(#1 \mid #2, #3)}}

\newcommand{\isEquiv}[3]{\ensuremath{{#1}\ {#3}\ {#2}}}
\renewcommand{\Inv}[2]{\ensuremath{\pmb{\Sigma}_{#1}(#2)}}
\renewcommand{\Inv}[2]{\ensuremath{{\Sigma}(#1, #2)}}
\renewcommand{\SubProbSenza}{\spaceFont{\$}}
\renewcommand{\SubProb}[1]{\SubProbSenza(#1)}
\newcommand{\BorelSenza}{\spaceFont{B}}
\renewcommand{\Gilt}[2][\phi]{\LinkeKlammer#1\RechteKlammer_{#2}}
\def\bS#1{\ensuremath{\BorelSenza\bigl(\SubProb{#1}\bigr)}}
\def\phi{\varphi}
\def\theta{\vartheta}
\def\wrd#1{{#1}^\star}
\def\QED{\ensuremath{\hfill\Box}}
\def\AE{\Psi}
\makeatletter
\newcommand{\@testOp}[2]{#1#2}
\newcommand{\posTest}[1][\phi]{\@testOp{#1}{?}}
\newcommand{\negTest}[1][\phi]{\@testOp{#1}{\text{\textquestiondown}}}
\makeatother

\begin{document}
\maketitle
\begin{abstract}
  Game logic is a dynamic modal logic which models strategic two person games; it contains propositional dynamic logic (PDL) as a fragment. We propose an interpretation of game logic based on stochastic effectivity functions. A definition of these functions is proposed, and some algebraic properties of effectivity functions such as congruences are investigated. The relationship to stochastic relations is characterized through a deduction system. Logical and behavioral equivalence of game models is investigated. Finally the completion of models receives some attention. 
\end{abstract}

\section{Introduction}
\label{sec:intro}

``Game logics describe general games through powers of players for forcing outcomes.'' This is the general description with which  van Benthem opens his paper~\cite{Benthem-LogicGames} on game logics. Models for situations to which such a logic may apply are general two person games of the kind described by Zermelo~\cite{Zermelo-Schach}, markets of commodities, the analysis of arguments between a proponent and a critic of a claim, termination of distributed systems and many more. Two players play against each other, each player working toward a winning situation. It is assumed that exactly one of the players will win the game, so draws are excluded. The players are considered to be equivalent, so that not one player dominates the other one, in fact, we will assume that the actions of player~$II$ can be modelled by the actions of player~$I$ simply by interchanging their r\^oles. Formally, the stage of our play is a set of states; given a state and a game, each player has in this state certain possibilities to force an outcome when playing this game, i.e., a set of states from which the next state may be selected. We do not give, however, policies which help the players in arriving at decisions which next state or even which set of possible next states to select. The mechanisms for these decisions are assumed to be outside the realm of the game.   

\paragraph{Effectivity Functions.}
Parikh~\cite{Parikh-Games1985}, and later Pauly~\cite{Pauly-CWI} propose interpreting game logic through a neighborhood model. Assign to each primitive game $g$ and each player $\{1, 2\}$ a neighborhood relation $N_g^{(i)} \subseteq S \times \PowerSet{S}\ (i = 1, 2)$ with the understanding that $s N_g^{(i)} X$ indicates player $i$ having a strategy in state $s$ to force a state in $X \subseteq S$. Here $S$ is the set of states over which the game is interpreted. The fact that $s N_g^{(i)} X$ is sometimes described by saying that player $i$ is effective for $X$ (with game $g$ in state $s$). It is desirable that $s N_g^{(i)} X$ and $X \subseteq X'$ imply $s N_g^{(i)} X'$ for all states $s$. We assume that the game is \emph{determined}, i.e., that exactly one of the players has a winning strategy. Thus $X \subseteq S$ is effective for player $I$ in state $s$ if and only if $S \setminus X$ is not effective for player $II$ in that state. Consequently, 
\begin{equation}
\label{intro:det}
s N_g^{(2)} X \Leftrightarrow\neg(s N_g^{(1)} S\setminus X), 
\end{equation}
which in turn implies that we only have to cater for player~$I$. We will omit the superscript from the neighborhood relation $N_g$. Define the map 
$
H_g: S \to \PowerSet{\PowerSet{S}}
$ 
upon setting
$
H_g(s) := \{X \subseteq S \mid s N_g X\},
$
then $H_g(s)$ is for all $s \in S$ an upper closed subset of $\PowerSet{S}$ from which relation $N_g$ can be recovered. This function is called the \emph{effectivity function} associated with relation $N_g$. From $N_g$ another map 
$
\widetilde{N_g}: \PowerSet{S}\to\PowerSet{S}
$
is obtained upon setting
$
\widetilde{N}_g(A) := \{s \in S \mid s N_g A\} = \{s \in S \mid A \in H_g(s)\}.
$
Thus state $s$ is an element of $\widetilde{N_g}(A)$ iff the first player has a strategy force the outcome $A$ when playing $g$ in $s$. The operations on games can be taken care of through this family of maps, e.g., one sets recursively
\begin{align*}
\widetilde{N}_{g_1\cup g_2}(A) & := \widetilde{N}_{g_1}(A)\cup \widetilde{N}_{g_2}(A),\\
\widetilde{N}_{g_1;g_2}(A) & := (\widetilde{N}_{g_1}\circ \widetilde{N}_{g_2})(A),\\
\widetilde{N}_{g^*} & := \bigcup_{n \geq 0} \widetilde{N}_{g^n}(A).
\end{align*}
with $g_1\cup g_2$ denoting the game which chooses from games $g_1$, $g_2$, the game $g_1;g_2$ plays $g_1$ first, then $g_2$, and $g^*$ is the indefinite iteration of game $g$; these operations will be discussed in greater detail in Section~\ref{sec:prog-games}. This refers only to player~$I$, player~$II$ is accommodated through $A \mapsto S\setminus N_g(S \setminus A)$ by~(\ref{intro:det}). The maps $\widetilde{N}_g$ serve in Parikhs's original paper as a basis for defining the semantics of game logic. It turns out to be convenient for the present paper, however, to use effectivity functions as maps to upper closed subsets, see Section~\ref{sec:effFnct}. 

The neighborhood relations used here are taken from the minimal models discussed in modal logics~\cite[Chapter
 7.1]{Chellas} or~\cite{Venema-Handbook}, serving as  basic mechanism for  models which are more general than Kripke models. The association of the effectivity functions sketched here to a very similar notion investigated in economics~\cite{Moulin-SocialChoice,Abdou-Keiding} is discussed in the survey paper~\cite[Section 2.3]{Hoe-Pauly}. Pauly~\cite[Section 6.3]{Pauly-CWI} discusses the important point of determinacy of games and relates it briefly to the discussion in set theory~\cite[Section 33]{Jech} or~\cite[Section 20]{Kechris}.

\paragraph{Motivation.}
The interpretation of game logic through various models is fairly well understood, but a stochastic interpretation is lacking. One way of doing this would be to generalize the stochastic interpretation of general modal logics~\cite{Panangaden-book,EED-CoalgLogic-Book} to game logics. Thus a model for this logics would be a stochastic Kripke model, assigning a stochastic relation to each primitive game and finding suitable constructions for composite games along the lines of~\cite{EED-PDL-TR}. This, however, would cover only part of game logics in the same way a non-deterministic Kripke models extended to game logics would only yield a partial understanding of these logics. Witness to this inadequacy is the observation that in Kripke models the choice operator distributes from the left and from the right over the composition operator. But it may be argued that left distributivity is not adequate in all situations, right distributivity, however, is. A stochastic interpretation of game logics requires consequently an extended formalism which covers and extends stochastic Kripke models. We  propose in this paper the use of stochastic effectivity functions. They model the players' behavior by assigning each primitive game and each state a portfolio, i.e., a set of distributions according to which the new state after playing the game in that state is distributed. The set of all possible portfolios for a state is originally given only for primitive games, it will be extended to composite ones; the set of portfolios has to satisfy certain measurability conditions for enabling this. The models so constructed should cover stochastic Kripke models as well, hence for example a probabilistic interpretation of PDL should arise as a special case. 
 
\paragraph{Related Work.}
Game logic was proposed originally by Parikh~\cite{Parikh-Games1985} who suggested neighborhood functions as a framework for its interpretation. This was later refined and extended by Pauly, partly together with Parikh~\cite{Pauly-CWI,Pauly-Parikh}. This has turned out a fruitful area of research: Venema~\cite{Venema-game-algs} gives a representation of game algebras, in which the equivalence of game algebras and board algebras is proved. A game algebra is a de Morgan lattice with two additional operations, one for passing control to the second player, the other one corresponding to the composition of games; this is syntactically oriented. A board algebra is semantically oriented by investigating suitable pairs of relations with operations corresponding to the game operations. The equivalence is in spirit quite close to Stone's Representation Theorem which relates Boolean algebras (a syntactic device) to fields of sets (a semantic device). Goranko~\cite{Goranko} adopts a similar algebraic approach; he gives a complete axiomatization of the identities for the basic game algebra which hold with respect to the game board semantics.  A first step toward a coalgebraic treatment is proposed in~\cite{EED-Game-Coalg} where it is shown that bisimulations may be treated as spans of morphisms for effectivity functions.

Propositional dynamic logics (PDL) is a fragment of game logic, its interpretation in terms of nondeterministic Kripke models is given in relational terms. The interesting point here is that this interpretation uses implicitly the power set monad through the monad's Kleisli construction. Another monad, viz., the Giry monad is used in Kozen's interpretation of PDL~\cite{Kozen-ProbPDL} through Kleisli morphisms and their composition, so does~\cite{EED-PDL-TR}. These papers are oriented toward the usual Kripke semantics for modal logics. The discussion on effectivity functions indicates that this approach is not general enough to game logics, and the use of monads does not appear to be helpful either.

There is a marked difference between the games we investigate here and the games which are investigated in set theory~\cite[Chapter~33]{Jech},~\cite[Section~12.3]{Jech-Choice},~\cite{Kechris,Herrlich-Choice}. The games investigated there are used as tools for investigating properties of objects through looking at \emph{strategies} the players might have. We do here  without explicit strategies when investigating the semantic properties of a modal logic, the modalities of which are derived from a game; the notion of a strategy as a map directing the action of one of the players is in our context fairly meaningless. Nevertheless we postulate in the present paper that the game is determined, yielding an interesting duality between the players.

\paragraph{Overview.}
Section~\ref{sec:prog-games} gives a syntax for games, formally introducing the operations on games. It defines also the modal logics which will be investigated. Because we will need some tools from measure theory, we provide for the reader's convenience a brief discussion of measurable spaces and some basic constructions. Here some material about measurable functions and invariant sets can be found as well. All this is happens in Section~\ref{sec:meas}. We will interpret game logic through a game model which is comprised of stochastic effectivity functions. These functions are introduced and investigated in Section~\ref{sec:effFnct}. Stochastic relations give rise to stochastic effectivity functions, we give sufficient and necessary conditions for an effectivity function to be generated by a stochastic relation, indicating a connection to deduction systems. We also introduce morphisms and congruences there, shedding some light on the algebraic properties of these functions. Game frames as a further step toward game models are introduced in Section~\ref{sec:gameFr-Trans}, and we show how games are used to transform sets of states, assigning each game a family of set transformations. We show that measurable sets are transformed into measurable sets, provided the underlying measurable space is closed under the Souslin operation, one of the basic operations in descriptive set theory. We investigate also frames that are generated from stochastic Kripke frames, and prove that sequential composition of games is fully distributive over angelic choice in these frames; this is in contrast to general frames where composition is only right distributive. After all these preparations Section~\ref{sec:game-models} introduces game models, the interpretation of formulas is defined and some properties are investigated, among others the observation that model morphisms are compatible with validity. The logic gives rise to an equivalence relation on the state space; the conditions under which this forms a congruence are investigated. This is a first step toward looking at the relationship of logical and behavioral equivalence of models, and is done for a class of models which satisfy the Frege condition. In Section~\ref{sec:test-op} we show also that the test operator, which has been neglected until now, can be integrated easily. Because the measurability of the validity sets requires stability under the Souslin operation, and because universally complete measurable spaces are closed under this operation, we discuss a technique for completing a model by completing the underlying space and extending the operations accordingly. This happens in Section~\ref{sec:complete}. Section~\ref{sec:concl} wraps it all up and indicates where further work could be done. 


\section{Programs and Games}
\label{sec:prog-games}
Game logic is a modal logic in which the modalities are given by games~\cite{Parikh-Games1985,Pauly-Parikh}. The formulas of the modal logic which we will consider are given through this grammar
\begin{equation}
\label{mod-grammar}
  \phi = \top~\mid~p~\mid~\phi_1\wedge\phi_2~\mid~\langle \tau \rangle_q \phi.
\end{equation}
Here $p\in\AE$ is a primitive formula, $q$ is a numerical value, and $\tau$ is a game. The logic is negation free, it has apart from conjunction a decorated modal operator. Intuitively, formula $\langle \tau \rangle_q \phi$ is true in state $s$ if playing game $\tau$ in state $s$ will result in a state in which formula $\phi$ holds with a probability greater than $q$.

Games in turn are given by this grammar
\begin{equation}
\label{game-grammar}
\tau ::= \gamma~\mid~\tau^d~\mid~\tau_1\cup\tau_2~\mid~\tau_1\cap\tau_2~\mid~\tau_1;\tau_2~\mid~\tau^*~\mid~\tau^\times
\end{equation}
with $\gamma\in\Gamma$, the set of atomic games. This models a two person game, one player is called \emph{Angel}, the other one \emph{Demon}. Angel plays against Demon. 
Games can be combined in different ways. If $\tau$ and $\tau'$ are games, $\tau;\tau'$ is the sequential composition of $\tau_1$ and $\tau_2$, so that plays $\tau$ first, then $\tau'$. In the game $\tau\cup\tau'$, Angel has the first move and decides whether $\tau$ or $\tau'$ is to be played, then the chosen game is played; $\tau\cup\tau'$ is called the \emph{angelic choice} between $\tau_1$ and $\tau_2$. Similarly, in $\tau\cap\tau'$ Demon has the first move and decides whether $\tau$ or $\tau'$ is to be played; accordingly, $\tau\cap\tau'$ is the \emph{demonic choice} between the games. In the game $\tau^*$, game $\tau$ is played repeatedly, until Angel decides to stop; it is not said in advance how many times the game is to be played, but it has to stop at some time; this is called \emph{angelic iteration}. Dually, Demon decides to stop for the game $\tau^\times$; this is called \emph{demonic iteration}. Finally, the rles of Angel and Demon are interchanged in the game $\tau^d$, so all decisions made by Demon are now being made by Angel, and vice versa. 

An important class of games is given by programs, which can be perceived as those games that are being played with one player only. Programs are given by this grammar
\begin{equation}
\label{prog-grammar}
\tau ::= \pi~\mid~\tau_1\cup\tau_2~\mid~\tau_1;\tau_2~\mid~\tau^*
\end{equation}
with $\pi\in\Pi$, the set of primitive programs. The corresponding logic is usually called \emph{propositional dynamic logic}, abbreviated as PDL. 

Thus programs can be combined through sequential composition and through the choice operator; we have also  indefinite iteration of programs. 

It is noted that in this version neither games nor programs have a test operator (which is usually provided both with game logic and with PDL). We will discuss the test operator, however, in Section~\ref{sec:test-op}.
\section{Measurable Spaces}
\label{sec:meas}

A measurable space $S$ is a pair comprised of a carrier set together with a $\sigma$-algebra on it; we denote the carrier set also by $S$ and the $\sigma$-algebra by $\Borel{S}$. The member of $\Borel{S}$ are usually called the \emph{measurable sets} of $S$. 
The set $\SubProb{S}$ of all subprobabilities is endowed with the weak-*-$\sigma$ algebra; this is the smallest $\sigma$-algebra which renders the evaluation $\mu \mapsto \mu(A)$ for each measurable set $A\in\Borel{S}$ measurable. It has as a base the sets
$
\{\bas{A}{\bowtie q} \mid A \in \Borel{S}, q \in \pReal\}
$
where
\begin{equation*}
\bas{A}{\bowtie q} := \{\mu\in\SubProb{S} \mid \mu(A) \bowtie q\} 
\end{equation*}
with $\bowtie$ one of the relations $\{<, \leq, \geq, >\}$. If necessary, we note the base space as well, writing down the set above as
$
\basS{S}{A}{\bowtie q}.
$

If $\mathcal{C}\subseteq \Borel{S}$ and $\mathcal{D} \subseteq \Borel{T}$ are sub-$\sigma$-algebras of the respective measurable sets, then a map $f: S \to T$ is called \emph{$\mathcal{C}$-$\mathcal{D}$-measurable} iff 
$
\InvBild{f}{\mathcal{D}} := \{\InvBild{f}{D}\mid D \in \mathcal{D}\} \subseteq \mathcal{C},
$
so the inverse image under $f$ of each set in $\mathcal{D}$ is a member of $\mathcal{C}$. If $f$ is $\Borel{S}$-$\Borel{T}$-measurable, $f$ is simply called \emph{measurable}. Now let $f: S \to T$ be a measurable map, then $\SubProb{f}: \SubProb{S}\to\SubProb{T}$ is defined through 
\begin{equation*}
\SubProb{f}(\mu)(B) := \mu(\InvBild{f}{B})
\end{equation*}
for $\mu\in\SubProb{S}, B \in \Borel{T}$. $\SubProb{f}$ is measurable as well, since both $\SubProb{S}$ and $\SubProb{T}$ carry the weak-*-$\sigma$ algebra, and because
$
\InvBild{\SubProb{f}}{\basS{T}{B}{\bowtie q}} = \basS{S}{\InvBild{f}{B}}{\bowtie q}
$
for $B \in \Borel{T}$. 

\BeginDefinition{stochRel}
A \emph{stochastic relation} $K: S \Trans T$ for the measurable spaces $S$ and $T$ is a measurable map $K: S \to \SubProb{S}$. 
\EndDefinition

A central notion for the present paper is that of a stochastic relation, which is sometimes also called a transition subprobability. 

Thus $K: S \Trans T$ is a stochastic relation iff
\begin{enumerate}
\item $K(s)$ is for each $s\in S$ a subprobability measure on the measurable space $T$,
\item the map $s \mapsto K(s)(C)$ is measurable for each Borel set $C \in T$.
\end{enumerate}
The first property is due to $K(s)$ being a member of $\SubProb{T}$, the second is implied by measurability, since it entails for each $C \in T$ and each $q \geq 0$ that 
\begin{equation*}
\{s \in S \mid K(s)(C) \bowtie q\} = \InvBild{K}{\bas{C}{\bowtie q}} \in \Borel{S}.
\end{equation*}
This is Giry's view~\cite{Giry}: $\SubProbSenza$ is an endofunctor on the category of all measurable spaces with measurable maps as morphisms, the functor is the functorial part of a monad which is sometimes called the \emph{Giry monad}. The  Kleisli morphisms for this monad are just the stochastic relations, for a discussion see, e.g.,~\cite{EED-CoalgLogic-Book}. 

A particularly easy structured stochastic relation is the \emph{Dirac relation} $D$ which assigns to each state the measure which gives it mass $1$: define for $s \in S$ and $A \in \Borel{S}$
\begin{equation}
\label{Dirac}
D(s)(A) := \delta_s(A) := 
\begin{cases}
1     & \text{if } s \in A \\
0     & \text{otherwise}.
\end{cases}
\end{equation} 
Then
\begin{equation*}
\{s \in S \mid D(s)(A) \leq q\} = 
\begin{cases}
S & \text{if } q \geq 1, \\
S\setminus A      & \text{if } 0 \leq q < 1, \\
\emptyset & \text{ otherwise}.
\end{cases}
\end{equation*}
Thus $s \mapsto D(s)(A)$ is measurable whenever $A \in \Borel{S}$, and $A \mapsto D(s)(A)$ defines a member of $\SubProb{S}$.

Given measurable spaces $S$ and $T$,  their  product $S \otimes T$ has the Cartesian product $S \times T$ as its carrier set, and is endowed with the  $\sigma$-algebra $\Borel{S}\otimes\Borel{T}$ which is initial with respect to the projections $\langle s, t\rangle \mapsto s$ and $\langle s, t\rangle \mapsto t$. Thus
\begin{equation*}
\Borel{S\otimes T} := \Borel{S}\otimes\Borel{T} = \sigma(\{B \times C \mid B \in \Borel{S}, C \in \Borel{T}\}),
\end{equation*} 
$\sigma(\mathcal{A})$ denoting the smallest $\sigma$-algebra containing $\mathcal{A}$, which is then called a \emph{generator} for its $\sigma$-algebra. Note that we overload the symbol $\otimes$ somewhat: on the level of measurable spaces, we denote the Cartesian product of the spaces, while we mean on the level of $\sigma$-algebras the product-$\sigma$-algebra, i.e., the smallest $\sigma$-algebra containing all measurable rectangles. Both notations are used interchangeably.

If $(T, \tau)$ is a topological space, then the $\sigma$-algebra $\sigma(\tau)$ is called the $\sigma$-algebra of \emph{Borel sets} of $T$. We assume that a topological space is endowed with its Borel sets, and do not distinguish between the topological space and its measurable offspring. A \emph{Polish space} is a second countable topological space which can be metrized with a complete metric (sometimes the measurable space associated with a Polish space is called a \emph{Standard Borel} space). 

Denote for  $A \subseteq S \times T$ the \emph{horizontal cut of $A$ at $t \in T$} by 
$
A_t := \{s \in S \mid \langle s, t\rangle \in A\}.
$
\medskip
Integration together with the cut provides us with a wealth of measurable maps which will be used extensively; its simplest form is given in this way:
\BeginLemma{meas-1}
Let  $B \in \Borel{S\otimes[0, 1]}$, then 
\begin{equation*}
\Lambda_B(\mu) := \int_0^1 \mu(B_r)\ dr
\end{equation*}
defines a measurable map $\SubProb{S}\to [0, 1]$.
\EndLemma

\BeginProof
Consider 
\begin{equation*}
\mathcal{C} := \{B \in \Borel{S\otimes[0, 1]} \mid \Lambda_B\text{ is measurable}\}.
\end{equation*}
$\mathcal{C}$ is closed under countable disjoint unions and under complementation, so it suffices to show that $\mathcal{C}$ contains a generator for $\Borel{S\otimes[0, 1]}$ which is closed under finite intersections. Then the assertion will follow from Proposition~\ref{pi-lambda}, the $\pi$-$\lambda$-Theorem. 

In fact, if $B = C \times V$ with $C\in\Borel{S}$ and $V\in\Borel{[0, 1]}$ is a measurable rectangle, we see that 
$
\Lambda_{C\times V}(\mu) = \mu(C)\cdot\lambda(V)
$
(with $\lambda$ as the Lebesgue measure on $[0, 1]$), so $\Lambda_{C\times V}$ is measurable. Since the measurable rectangle generate the product $\sigma$-algebra, and since the set of all rectangles is closed under finite intersections, the assertion follows. 
\EndProof
 
This has as an easy consequence

\BeginCorollary{meas-2}
Let  $B \in \Borel{S\otimes[0, 1]}$, then 
\begin{equation*}
\{\langle \mu, q\rangle \in\SubProb{S}\times[0, 1]\mid  \int_0^1\mu(B_r)\ dr \bowtie q\}
\end{equation*}
defines a measurable subset of $\SubProb{S}\times[0, 1]$.
\EndCorollary

\BeginProof
Because the map
$
\mu \mapsto \int_0^1\mu(B_r)\ dr
$
is measurable, the assertion follows from Choquet's Theorem (Theorem~\ref{Choquet}, (\ref{Choquet-1})). 
\EndProof

\medskip

\subsection{Tame Relations}
\label{sec:tame-rel}

Given an equivalence relation $\rho$ on $S$, call a subset $U\subseteq S$ \emph{$\rho$-invariant} iff $U$ is the union of $\rho$-equivalence classes, or, equivalently, iff $s \in U$ and $\isEquiv{s}{s'}{\rho}$ together imply $s'\in U$. Denote by $\Inv{\rho}{S}$ the $\sigma$-algebra of invariant measurable subsets of $S$, thus
\begin{equation*}
\Inv{\rho}{S} := \{U \in \Borel{S} \mid U\text{ is $\rho$-invariant}\}.
\end{equation*}
Sometimes the measurable space $(S, \Inv{\rho}{S})$ is denoted by $\Inv{\rho}{S}$ as well. 

Conversely,  a $\sigma$-algebra $\mathcal{A}\subseteq\Borel{S}$ on the measurable space $S$ induces an equivalence relation $\rho_{\mathcal{A}}$ upon setting 
\begin{equation}
\label{det-equiv}
\isEquiv{s}{s'}{\rho_{\mathcal{A}}} :\Longleftrightarrow [\forall A \in \mathcal{A}_0: s \in A \text{ iff } s'\in A]
\end{equation}
for some generator $\mathcal{A}_0$ of $\mathcal{A}$. It is easy to see that each element of $\mathcal{A}$ is $\rho_{\mathcal{A}}$-invariant. But we do not have necessarily 
$
\Inv{\rho_{\mathcal{A}}}{S} = \mathcal{A}:
$
Take for example $S = (\Real, \Borel{\Real})$, where the measurable sets are the Borel sets of the usual topology, and take $\mathcal{Y}$ as the countable-cocountable sub-$\sigma$-algebra of $\Borel{\Real}$, then $\rho_{\mathcal{Y}}$ is the identity, and 
$
\Inv{\rho_{\mathcal{Y}}}{S} = \Borel{\Real}.
$
\BeginDefinition{exact}
Given an equivalence relation $\rho$ and a subset $\mathcal{A}\subseteq\Borel{S}$ of the measurable sets of $S$. Call \emph{$\rho$ exact with $\mathcal{A}$} iff 
$
\Inv{\rho}{S} = \sigma(\mathcal{A}).
$
\EndDefinition
Thus $\mathcal{A}$ generates exactly the invariant measurable sets of $\rho$; it is easy to see that $\mathcal{A}$ determines $\rho$ as in~(\ref{det-equiv}). Taking $\mathcal{Y}$ as above, we see that $\rho_{\mathcal{Y}}$ is not exact with $\mathcal{Y}$; it is, exact, however, with the open sets $\mathcal{G}$ or the intervals $\mathcal{I}$ of $\Real$, because 
$
\Inv{\rho_{\mathcal{Y}}}{S} = \Borel{\Real} = \sigma(\mathcal{G}) = \sigma(\mathcal{I}).
$

The set $\Faktor{S}{\rho}$ of all equivalence classes is endowed with the final $\sigma$-algebra with respect to the factor map 
$
\fMap{\rho}: s \mapsto \Klasse{s}{\rho},
$
i.e., the largest $\sigma$-algebra rendering $\fMap{\rho}$ measurable. Hence 
\begin{equation*}
\Borel{\Faktor{S}{\rho}} = \{C \subseteq \Faktor{S}{\rho} \mid \InvBild{\fMap{\rho}}{C} \in \Borel{S}\}.
\end{equation*}
It follows that $\Bild{\fMap{\rho}}{B}\in\Borel{\Faktor{S}{\rho}}$ whenever $B \in \Inv{\rho}{S}$, because 
$
B = \InvBild{\fMap{\rho}}{\Bild{\fMap{\rho}}{B}}
$
on account of the invariance of $B$. Accordingly, call a measurable map $f: S \to T$ \emph{final} iff $\Borel{T}$ is the final $\sigma$-algebra with respect to $f$ (and $\Borel{S}$). Thus if $f$ is final, we may conclude from $\InvBild{f}{C}\in\Borel{S}$ that $C \in \Borel{T}$, and 
$ 
\Borel{T} = \{\Bild{f}{A} \mid A \in \Inv{\Kern{f}}{S}\}.
$ 
Here 
\begin{equation*}
\Kern{f} := \{\langle s, s'\rangle \mid s, s'\in S, f(s) = f(s')\}
\end{equation*}
is the \emph{kernel} of $f$.

The first part of the following statement is obvious.
\BeginLemma{kerf-is-exact}
Let $f: S \to T$ be a measurable map, then $\Kern{f}$ is exact with $\Inv{\Kern{f}}{S}$. If $f$ is final, then 
$
\InvBild{f}{\Borel{T}} = \Inv{\Kern{f}}{S}.
$
\EndLemma

{
\def\fKf{\fMap{\Kern{f}}}
\def\xKf{\Faktor{S}{\Kern{f}}}
\BeginProof
Since $f$ is measurable and $\InvBild{f}{B}$ is an invariant set, it is clear that
$
\InvBild{f}{\Borel{T}} \subseteq \Inv{\Kern{f}}{S}
$
holds. The measurable map $f: S \to T$ can be decomposed as 
$
f = \widetilde{f}\circ\fKf
$
with 
$
\widetilde{f}: \xKf\to T
$
injective, hence $\widetilde{f}^{-1}$ is surjective. Because $\xKf$ has the final $\sigma$-algebra with respect to $\fKf$, $\widetilde{f}$ is measurable. Because $f$ is final, we conclude that 
$
\InvBild{\widetilde{f}}{B}\in\Borel{\xKf}
$
entails 
$
B \in \Borel{T}.
$

Now let $D \in \Inv{\Kern{f}}{S}$, then $\Bild{\fMap{\Kern{f}}}{D}\in\Borel{\Faktor{S}{\Kern{f}}}$. Hence can find $E \subseteq T$ with 
$
\Bild{\fMap{\Kern{f}}}{D} = \InvBild{\widetilde{f}}{E},
$
and because
$
D = \InvBild{\fMap{\Kern{f}}}{\Bild{\fMap{\Kern{f}}}{D}}
$
on account of $D$ being $\fMap{\Kern{f}}$-invariant we conclude 
$
D = \InvBild{f}{E},
$ 
so that $E \in \Borel{T}$. This implies
$
\InvBild{f}{\Borel{T}} \supseteq \Inv{\Kern{f}}{S}.
$
\EndProof
}

\medskip

If $\rho$ is exact with $\mathcal{A}$, we have  a handle on the elements of $\Borel{\Faktor{S}{\rho}}$, albeit in a special situation. The characterization below is very similar to Corollary 2.6.5 in~\cite{EED-CoalgLogic-Book}, which deals with validity sets of the formulas of a negation free logic which is closed under finite conjunctions. The proof easily carries over  to the situation at hand. 

\BeginLemma{exact-generate}
Let $\rho$ be exact with $\mathcal{A}$, and assume that $\mathcal{A}$ is closed under finite intersections. Then we have
\begin{enumerate}
\item $\Inv{\rho}{S} = \sigma(\mathcal{A})$,
\item $\Borel{\Faktor{S}{\rho}} = \sigma(\{B \subseteq \Faktor{S}{\rho} \mid \InvBild{\fMap{\rho}}{B} \in \mathcal{A}\}).$
\QED
\end{enumerate}
\EndLemma

We need a slightly stronger condition on the equivalence relations we are dealing with, because we need to consider reals as well. Define for this the equivalence relation $\rho\times\Delta$ on $S \times [0, 1]$ upon setting
\begin{equation*}
\isEquiv{\langle s, q\rangle}{\langle s', q'\rangle}{(\rho\times\Delta)} \text{ iff } \isEquiv{s}{s'}{\rho}\text{ and } q = q'.
\end{equation*}

Let us call an equivalence relation $\rho$ tame if the invariant sets of $\rho\times\Delta$ behave well. In descriptive set theory, countably generated equivalence relations on a Polish space are sometimes called tame (they are called \emph{smooth} in~\cite{EED-CoalgLogic-Book}); the behavior of the present tame relations is modelled after them.
 
\BeginDefinition{tame}
Call an equivalence relation $\rho$ on the measurable space $S$ \emph{tame} iff these conditions hold
\begin{enumerate}
\item $\rho$ is exact with some $\mathcal{A}\subseteq \Borel{S}$,
\item $\rho\times\Delta$ is exact with $\{A \times I \mid A\in \mathcal{A}, I\subseteq [0, 1]\text{ is an interval}\}$
\end{enumerate}
\EndDefinition

Thus, if we know that $\Inv{\rho}{S} = \sigma(\mathcal{A})$ for tame $\rho$, then we may identify a generator for $\sigma$-algebra $\Inv{\rho\times\Delta}{S\otimes[0, 1]}$ through the rectangles composed from elements of $\mathcal{A}$ and intervals in the unit interval.  Consequently, dealing with the invariant sets for $\rho\times\Delta$ becomes more practical through tameness:
\BeginLemma{is-tame}
Let $\rho\times\Delta$ be tame, then 
$
\Inv{\rho\times\Delta}{S\otimes[0, 1]} = \Inv{\rho}{S}\otimes\Borel{[0, 1]}.
$
\EndLemma

\BeginProof
It is easy to see that $\Inv{\rho}{S}\otimes\Borel{[0, 1]}$ is a subset of  $\Inv{\rho\times\Delta}{S\otimes[0, 1]}$, because each measurable rectangle $B \times G \in \Inv{\rho}{S}\otimes\Borel{[0, 1]}$ is a $\rho\times\Delta$-invariant measurable set, hence $B\times G \in \Inv{\rho\times\Delta}{S\otimes[0, 1]}$. Thus the $\sigma$-algebra generated by these sets is contained in the latter as well. Conversely, we infer from tameness that
\begin{align*}
\Inv{\rho\times\Delta}{S\otimes[0, 1]}
& = 
\sigma(\{A \times I\mid A\in \mathcal{A}, I\text{ is an interval}\})\\
& \subseteq 
\sigma(\{A \times B \mid A \in \sigma(\mathcal{A}), B \in \Borel{[0, 1]}\})\\
& = 
\Inv{\rho}{S}\otimes\Borel{[0, 1]}.
\end{align*}
\EndProof

Thus we can characterize the factor space with respect to $\rho\times\Delta$ easily:

\BeginCorollary{cor-is-tame}
Assume that $\rho$ is tame, then 
$
\Faktor{(S\otimes [0, 1])}{\rho\times\Delta}
$
and
$
\Faktor{S}{\rho}\otimes[0, 1]
$
are Borel isomorphic.
\EndCorollary

\BeginProof
It is not difficult to see that 
\begin{equation*}
\tau: 
\begin{cases}
\Faktor{(S\otimes [0, 1])}{\rho\times\Delta} & \to \Faktor{S}{\rho}\otimes[0, 1]\\
\Klasse{\langle s, t\rangle}{\rho\times\Delta}&\mapsto \langle \Klasse{s}{\rho}, t\rangle
\end{cases}
\end{equation*}
defines a measurable map. 

For establishing that $\tau^{-1}$ is measurable, one notes first that $B \in \Inv{\rho}{S}\otimes\Borel{[0, 1]}$ implies 
$
\Bild{(\fMap{\rho}\times id)}{B} \in \Borel{\Faktor{S}{\rho}\otimes[0, 1]}.
$
This is so because the set of all $B$ for which the assertion is true is closed under complementation and countable disjoint unions, and it contains all measurable rectangles, so the assertion is established by Theorem~\ref{pi-lambda}.

Now take $H \in \Borel{\Faktor{(S\otimes[0, 1])}{\rho\times\Delta}}$, then 
\begin{equation*}
\InvBild{\fMap{\rho\times\Delta}}{H} \in \Inv{\rho\times\Delta}{S\otimes[0, 1]} = \Inv{\rho}{S}\otimes\Borel{[0, 1]},
\end{equation*}
since $\rho$ is tame. Hence 
$
\InvBild{(\fMap{\rho}\times id)}{\Bild{\tau}{H}}\in\Inv{\rho}{S}\otimes\Borel{[0, 1]},
$
so the assertion follows from
\begin{equation*}
\Bild{\tau}{H} = \Bild{(\fMap{\rho}\times id)}{\InvBild{(\fMap{\rho}\times id)}{\Bild{\tau}{H}}}.
\end{equation*}
\EndProof

We will deal with tame relations when we investigate congruences for game models. Specifically we want to know under which circumstances the equivalence related to game logic will be a congruence for the underlying model.


\subsection{Some Indispensable Tools}
\label{sec:indispensable}

We post here for the reader's convenience some measure theoretic tools which will be used all over.

\paragraph{The $\pi$-$\lambda$-Theorem.}
This technical tool is most useful when it comes to determine the $\sigma$-algebra generated by a family of sets. 

\BeginProposition{pi-lambda}
Let $\mathcal{A}$ be a family of subsets of a set $X$ that is closed under finite intersections. Then $\sigma(\mathcal{A})$ is the smallest family of subsets containing $\mathcal{A}$ which is closed under complementation and countable disjoint unions. 
\QED
\EndProposition

\paragraph{The Souslin Operation.}
$\wrd{V}$ denotes for a set $V$ the set of all finite words with letters from $V$ including the empty string $\epsilon$. Let 
$
\{A_s\mid s\in \wrd{\Nat}\}
$
be a collection of subsets of a set $X$ indexed by all finite sequences of natural numbers (a \emph{Souslin scheme}), then the \emph{Souslin operation $\mathfrak{A}$} on this collection is defined as
\begin{equation}
\label{Souslin}
\mathfrak{A}\bigl(\{A_s\mid s\in \wrd{\Nat}\}\bigr) := \bigcup_{\alpha\in\Nat^{\Nat}}\bigcap_{n\in\Nat} A_{\alpha|n},
\end{equation}
where $\alpha|n\in\wrd{\Nat}$ is just the word composed from the first $n$ letters of the sequence $\alpha$. This operation is intimately connected with the theory of analytic sets~\cite{Kuratowski-Mostowski,Jech,Bogachev,Kechris}. We obtain from~\cite[Proposition 1.10.5]{Bogachev}:

\BeginProposition{closed-under-Souslin}
If $S$ is a universally complete measurable space, then $\Borel{S}$ is closed under the operation $\mathfrak{A}$.
\QED
\EndProposition

\paragraph{Choquet's Representation.}
The following condition on product measurability and an associated integral representation attributed to Choquet is used~\cite[Corollary 3.4.3]{Bogachev}.
\BeginTheorem{Choquet}
Let $f: X \to \pReal$ be measurable and bounded, then
\begin{equation}
\label{Choquet-1}
C_{\bowtie}(f) := \{\langle x, r\rangle \in X \times \pReal \mid f(x) \bowtie r\} \in \Borel{X\otimes\pReal}.
\end{equation}
If $\mu$ is a $\sigma$-finite measure on $\Borel{X}$, then
\begin{equation}
\label{Choquet-2}
\int_X f(x)\ \mu(dx)  = \int_0^\infty \mu(\{x \in X \mid f(x) > t\})\ dt.
\end{equation}
\QED
\EndTheorem
The set 
$
C_>(f) = \{\langle x, r\rangle \in X \times \pReal \mid 0 \leq r < f(x)\}
$
may be visualized as the area between the $x$-axis and the graph of $f$. Hence formula~(\ref{Choquet-2}) specializes to the Riemann integral, if $f: \pReal\to\pReal$ is Riemann integrable.  

\section{Effectivity Functions}
\label{sec:effFnct}

For an interpretation of game logic, Parikh~\cite{Pauly-Parikh} and later Parikh and Pauly~\cite{Parikh-Games1985} use effectivity functions which are closely related to neighborhood relations~\cite{Pauly-CWI}. 
When constructing a probabilistic interpretation, we will take distributions over the state space into account ---~thus, rather than working with states directly, we will work with probabilities over them. It may well be that some information gets lost, so we choose to work with subprobabilities rather than probabilities. Taking into account that Angel may be able to bring about a specific distribution of the new states when playing $\gamma$ in state $s$, we propose that we model Angel's effectivity by a set of distributions (this is remotely similar to the idea of gambling houses in~\cite{Dubins+Savage}). For example,  Angel may have a strategy for achieving a normal distribution $\mathcal{N}(s, \sigma^2)$ centered at $s\in\Real$ such that the standard distribution varies in an interval $I$, yielding $\{\mathcal{N}(s, \sigma^2) \mid \sigma\in I\}$ as a set of distributions effective for Angel in that situation. 

But we  cannot do with just arbitrary subsets of the set of all subprobabilities on state space $S$. We want also to characterize possible outcomes, i.e., sets of distributions over the state space for composite games. This means that we will want to average over intermediate states, which in turn requires measurability of the functions involved. Hence we require measurable sets of subprobabilities as possible outcomes. We also impose a condition on measurability on the interplay between distributions on states and reals for measuring the probabilities of sets of states. This leads to the definition of a stochastic effectivity function. 

Modeling all this requires some preparations by fixing the range of a stochastic effectivity function through a suitable functor.  Put for a measurable space $S$ 
\begin{equation*}
\Vau{S} := \{V \subseteq \Borel{\SubProb{S}} \mid V \text{ is upward closed}\}
\end{equation*} 
thus if $V \in \Vau{S}$, then $A \in V$ and $A \subseteq B$ together imply $B\in V$. A measurable map $f: S \to T$ induces a map $\Vau{f}: \Vau{S}\to\Vau{T}$ upon setting
\begin{equation*}
\Vau{f}(V) := \{W \in \Borel{\SubProb{T}} \mid \InvBild{\SubProb{f}}{W}\in V\}
\end{equation*}
for $V \in \Vau{S}$, then clearly $\Vau{f}(V)\in\Vau{T}$. 

Note that $\Vau{S}$ has not been equipped with a $\sigma$-algebra, so the usual notion of measurability between measurable spaces cannot be applied. In particular, $\VauSenza$ is not an endofunctor on the category of measurable spaces. We will not discuss functorial aspects of $\VauSenza$ here. 

We need, however, some measurability properties for dealing with the composition of distributions when discussing composite games. Let $H \in \Borel{\SubProb{S} \otimes [0, 1]}$ be a measurable subset of $\SubProb{S}\times[0, 1]$ indicating a quantitative assessment of subprobabilities (a typical example could be 
$
\{\langle \mu, q\rangle \mid \mu \in \bas{A}{> q}, 0 \leq q \leq 1\}
$
for some $A \in \Borel{S}$). Fix some real $q$ and consider the set $H_q := \{\mu \mid \langle \mu, q\rangle \in H\}$ of all measures evaluated through $q$. We ask for all states $s$ such that this set is effective for $s$. They should come from a measurable subset of $S$. It turns out that this is not enough, we also require the real components being captured through a measurable set as well --- after all, the real component will be used to be averaged, i.e., integrated, over later on, so it should behave decently. This idea is captured in the following definition.

\BeginDefinition{t-measurability}
Call a map $P: S \to \Vau{S}$ \emph{t-measurable} iff 
$
\{\langle s, q\rangle \mid H_q\in P(s)\} \in \Borel{S\otimes[0, 1]}
$
whenever 
$ 
H \in \Borel{\SubProb{S} \otimes [0, 1]}. 
$ 
\EndDefinition

Summarizing, we are led to the notion of a stochastic effectivity function.

\BeginDefinition{effFnct}
A \emph{stochastic effectivity function} $P$ on a measurable space $S$ is a t-measurable map $P\to\Vau{S}$.
\EndDefinition

In order to distinguish between sets of states and sets of state distributions we call the latter ones \emph{portfolios}; thus $P(s)$ is a set of measurable portfolios. This will render some discussions below easier. 

A stochastic effectivity function between measurable spaces $S$ and $T$ could be defined in a similar way, but this added generality is not of interest in the present context.

The following technical statement will be helpful later on; it shows that t-measurability is preserved by averaging over reals, a property which will be of vital importance for the interpretation of formulas in game logic.

\BeginCorollary{meas-3}
Let $P: S \to \Vau{S}$ be a stochastic effectivity function, and assume
$
B \in \Borel{S\otimes[0, 1]}.
$
Put 
\begin{equation*}
B' := \{\langle\mu, q\rangle \in \SubProb{S}\times[0, 1]\mid \int_0^1\mu(B_r)\ dr \bowtie q\}.
\end{equation*}
Then
$
\{\langle s, q\rangle \in S \times [0, 1] \mid B'_q \in P(s)\} \in \Borel{S\otimes[0, 1]}.
$
\EndCorollary

\BeginProof
$B'$ is a measurable subset of $\SubProb{S}\times[0, 1]$ by Corollary~\ref{meas-2}, so the assertion follows from t-measurability.
\EndProof

The relationship of stochastic relations and stochastic effectivity functions is of great interest, because we will later on discuss stochastic Kripke models and general models for game logic. 

\subsection{Effectivity Functions vs. Stochastic Relations}
\label{sec:eff_vs_stoc}

Each stochastic relation $K: S \Trans S$ yields a stochastic effectivity function $P_K$ in a natural way upon setting
\begin{equation}
\label{stoch-to-eff}
P_K(s) := \{A \in \Borel{\SubProb{S}} \mid K(s)\in A\}.
\end{equation} 

Thus a portfolio in $P_K(s)$ is a measurable subset of $\SubProb{S}$ which contains $K(s)$. We observe

\BeginLemma{pk-is-t-measurable}
$P_K: S \to \Vau{S}$ is t-measurable, whenever $K: S \Trans S$ is a stochastic relation.
\EndLemma

\BeginProof
Clearly, $P_K(s)$ is upward closed for each $s \in S$. 
Put 
$
T_H := \{\langle s, q\rangle \mid H_q\in P_K(s)\}
$ 
for $H \subseteq \SubProb{S}\times [0, 1]$. Thus 
\begin{equation*}
\langle s, q\rangle \in T_H
\Leftrightarrow
K(s)\in H_q 
\Leftrightarrow
\langle K(s), q\rangle \in H
\Leftrightarrow
\langle s, q\rangle \in \InvBild{(K\times id_{[0, 1]})}{H}.
\end{equation*}
Because 
$
K\times id_{[0, 1]}: S \times [0, 1] \to \SubProb{S}\times [0, 1]
$
is a measurable function, $H \in \Borel{\SubProb{S}\otimes[0, 1]}$
implies $T_H\in\Borel{S\otimes[0, 1]}$. Hence $P_K$ is t-measurable. 
\EndProof

The following example will be of use later on. It shows that we have always a stochastic effectivity function, albeit a fairly trivial one.

\BeginExample{dirac}
Let $D: S \Trans S$ be the Dirac relation, cf.~(\ref{Dirac}), then 
$
I_D := P_D
$
defines an effectivity function, the \emph{Dirac effectivity function}. Consequently we have
$
W \in I_D(s) 
$
iff 
$
\delta_s \in W
$
for $W \in \Borel{\SubProb{S}}.$ This is akin to assigning each element of a set the ultra filter based on it.
\EndExample

The Dirac effectivity function will be useful for chacaterizing the effect of the empty game $\epsilon$, it will also help in modelling the effects of the test games $\posTest$ and $\negTest$ associated with formula $\phi$.

\medskip

We will use the construction~(\ref{stoch-to-eff}) for generating a model from a stochastic Kripke model, indicating that the models considered here are more general that Kripke models. The converse construction is of course of interest as well: Given a model, can we determine whether or not it comes from a Kripke model? This boils down to the question under which conditions a stochastic effectivity function is generated through a stochastic relation. We will deal with this problem now.

The tools for investigating the converse to Lemma~\ref{pk-is-t-measurable} come from the investigation of deduction systems for probabilistic logics. In fact, we are given a set of portfolios and want to know under which conditions this set is generated from a single subprobability. The situation is roughly similar to the one observed with deduction systems, where a set of formulas is given, and one wants to know whether this set can be constructed as valid under a suitable model. Because of the similarity, we may take (probably more than only) some inspiration from the work on deduction systems, and we adapt here the approach proposed by R. Goldblatt~\cite{Goldblatt-Deduction}. Goldblatt works with formulas while we are interested foremost in families of sets; this permits a technically somewhat lighter approach in the present scenario.

We  first have a look at a relation $R \subseteq [0, 1] \times \Borel{S}$ which models bounding probabilities from below. Intuitively, $\langle r, A\rangle\in R$ is intended to characterize the set $\bas{A}{\geq r}$.

\BeginDefinition{R-binding}
$R \subseteq [0, 1] \times \Borel{S}$ is called a \emph{characteristic relation on $S$} iff these conditions are satisfied
\begin{align*}
\text{\ding{172}}\ &\frac{\langle r, A\rangle\in R, A \subseteq B}{\langle r, B\rangle\in R} &
\text{\ding{173}}\  &\frac{\langle r, A\rangle\in R, r \geq s}{\langle s, A\rangle\in R}\\
\text{\ding{174}}\  &\frac{\langle r, A\rangle\notin R, \langle s, B\rangle\notin R, r + s \leq 1}{\langle r+s, A\cup B\rangle\notin R}&
\text{\ding{175}}\   &\frac{\langle r, A\cup B\rangle\in R, \langle s,A \cup (S\setminus B)\rangle\in R, r + s \leq 1}{\langle r+s, A\rangle\in R}\\
\text{\ding{176}}\   &\frac{\langle r, A\rangle\in R, r + s > 1}{\langle s, S \setminus A\rangle\notin R}&
\text{\ding{177}}\   &\frac{\langle r, \emptyset\rangle\in R}{r = 0}\\
\text{\ding{178}}\  &\frac{A_1 \supseteq A_2 \supseteq \dots, \forall n \in \Nat: \langle r, A_n\rangle\in R}{\langle r, \bigcap_{n\geq 1} A_n\rangle \in R}
\end{align*}
\EndDefinition
The conditions $\text{\ding{172}}$ and $\text{\ding{173}}$ make sure that bounding from below is monotone both in its numeric and in its set valued component. By $\text{\ding{174}}$ and $\text{\ding{175}}$ we cater for sub- and superadditivity of the characteristic relation, condition $\text{\ding{177}}$ sees to the fact that the probability for the impossible event cannot be bounded from below but through $0$, and finally $\text{\ding{178}}$ makes sure that if the members of a decreasing sequence of sets are uniformly bounded below, then so is its intersection. These conditions are adapted from the S-axioms for T-deduction systems in~\cite[Section 4]{Goldblatt-Deduction}. An exception is \ding{178} which is weaker than the Countable Additivity Rule in~\cite[Definition 4.4]{Goldblatt-Deduction}; we do not need a rule as strong as the latter one because we work with sets, hence we can deal with descending chains of sets directly. 

We show that each characteristic relation defines a subprobability measure; the proof follows \emph{mutatis mutandis} the proof of~\cite[Theorem 5.4]{Goldblatt-Deduction}.

\BeginProposition{def-measure}
Let $R \subseteq [0, 1] \times \Borel{S}$ be a \emph{characteristic relation on $S$}, and define for $A\in\Borel{S}$
\begin{equation*}
\mu_R(A) := \sup\{r \in [0, 1] \mid \langle r, A\rangle\in R\}.
\end{equation*} 
Then $\mu_R$ is a subprobability measure on $\Borel{S}$. 
\EndProposition

\BeginProof
1. 
$\text{\ding{177}}$ implies that $\mu_R(\emptyset) = 0$, and $\mu_R$ is monotone because of $\text{\ding{172}}$. It is also clear that $\mu_R(S) \leq 1$. We obtain from $\text{\ding{173}}$ that $\langle s, A\rangle\notin R$, whenever $s \geq r$ with $\langle r, A\rangle\notin R$. 

2. 
Let $A_1, A_2\in\Borel{S}$ be arbitrary. Then 
\begin{equation*}
\mu_R(A_1\cup A_2) \leq \mu_R(A_1) + \mu_R(A_2).
\end{equation*}
In fact, if 
$
\mu_R(A_1) + \mu_R(A_2) < q_1 + q_2 \leq \mu_R(A_1\cup A_2)
$
with 
$
\mu_R(A_i) < q_i\ (i = 1, 2),
$
then 
$
\langle q_i, A_i\rangle\notin R 
$
for $i = 1, 2$. Because $q_1 + q_2 \leq 1$, we obtain from $\text{\ding{174}}$ that 
$
\langle q_1 + q_2, A_1 \cup A_2\rangle\notin R.
$
By $\text{\ding{173}}$ this yields 
$
\mu_R(a_1\cup A_2) < q_1 + q_2,
$
contradicting the assumption. 

3. 
If $A_1$ and $A_2$ are disjoint, we observe first that 
$
\mu_R(A_1) + \mu_R(A_2) \leq 1.
$
Assume otherwise that we can find $q_i \leq \mu_R(A_i)$ for $i = 1, 2$ with $q_1 + q_2 > 1$. Because $\langle q_1, A_1\rangle\in R$ we conclude from $\text{\ding{176}}$ that 
$
\langle q_2, S\setminus A_2\rangle\notin R,
$
hence $\langle q_2, A_2\rangle\notin R$ by $\text{\ding{172}}$, contradicting $q_2 \leq \mu_R(A_2)$.  

This implies that 
\begin{equation*}
\mu_R(A_1) + \mu_R(A_2) \leq \mu_R(A_1) + \mu_R(A_2).
\end{equation*}
Assuming this to be false, we find $q_1 \leq \mu_R(A_1), q_2 \leq \mu_R(A_2)$ with
\begin{equation*}
\mu_R(A_1 \cup A_2) < q_1 + q_2 \leq \mu_R(A_1) + \mu_R(A_2).
\end{equation*}
Because $\langle q_1, A_1\rangle\in R$, we find 
$
\langle q_1, (A_1\cup A_2)\cap A_1\rangle\in R, 
$
because $\langle q_2, A_2\rangle\in R$ we see that
$
\langle q_2, (A_1\cup A_2)\cap (S\setminus A_1)\rangle\in R 
$
(note that $(A_1\cup A_2)\cap A_1 = A_1$ and $(A_1\cup A_2)\cap (S\setminus A_1) = A_2$, since $A_1\cap A_2 = \emptyset$). From $\text{\ding{175}}$ we infer that
$
\langle q_1 + q_2, A_1 \cup A_2\rangle\in R,
$
so that 
$
q_1 + q_2 \leq \mu_R(A_1 \cup A_2),
$
which is a contradiction. 

Thus we have shown that $\mu_R$ is additive.

4.
From $\text{\ding{178}}$ it is obvious that 
\begin{equation*}
\mu_R(A) = \inf_{n\in\Nat} \mu_R(A_n),
\end{equation*}
whenever 
$
A = \bigcap_{n\in\Nat} A_n
$
for the decreasing sequence $\Folge{A}$ in $\Borel{S}$. 
\EndProof

We  relate $Q\in\Vau{S}$ to the characteristic relation $R$ on $S$ by comparing $\bas{A}{\geq q}\in Q$ with $\langle q, A\rangle\in R$ by imposing a syntactic and a semantic condition. They will be shown equivalent. 

\BeginDefinition{satisf}
$Q\in\Vau{S}$ is said to \emph{satisfy} the characteristic relation $R$ on $S$ ($Q \vdash R$) iff we have 
\begin{equation*}
\langle q, A\rangle\in R \Leftrightarrow \bas{A}{\geq q}\in Q
\end{equation*}
for any $q\in [0, 1]$ and any $A\in\Borel{S}$.
\EndDefinition

This is a syntactic notion. Its semantic counterpart reads like this:

\BeginDefinition{implem}
$Q$ is said to \emph{implement} $\mu\in\SubProb{S}$  iff 
\begin{equation*}
\mu(A) \geq q \Leftrightarrow \bas{A}{\geq q}\in Q
\end{equation*}
for any $q\in [0, 1]$ and any $A\in\Borel{S}$. We write this as $Q \models\mu$.
\EndDefinition

Note that $Q\models \mu$ and $Q \models \mu'$ implies 
\begin{equation*}
\forall A \in \Borel{S}\forall q \geq 0: \mu(A) \geq q \Leftrightarrow \mu'(A) \geq q.
\end{equation*}
Consequently, $\mu = \mu'$, so that the measure implemented by $Q$ is uniquely determined. 

We will show now that syntactic and semantic issues are equivalent: $Q$ satisfies a characteristic relation if and only if it implements the corresponding measure. This will be used in a moment for a characterization of those game frames which are generated from Kripke frames.

\BeginProposition{satisf=implem}
$Q \vdash R$ iff $Q \models \mu_R$.
\EndProposition

\BeginProof
``$Q \vdash R~\Rightarrow~Q \models \mu_R$'':
Assume that $Q \vdash R$ holds. It is then immediate that $\mu_R(A) \geq r$ iff 
$\bas{A}{\geq r} \in Q$.

``$Q \models \mu_R~\Rightarrow~Q \vdash R$'':
If $Q \models \mu_R$ for relation $R \subseteq [0, 1] \times \Borel{S}$, we show that the conditions given in Definition~\ref{R-binding}  are satisfied. 
\begin{enumerate}
\item  Let $\bas{A}{\geq r}\in Q$ and $A \subseteq B$, thus $\mu_R(A) \geq r$, hence $\mu_R(B) \geq r$, which in turn implies $\bas{B}{\geq r}\in Q$. Hence $\text{\ding{172}}$ holds. $\text{\ding{173}}$ is established similarly.
\item If $\mu_R(A) < r$ and $\mu_R(B) < s$ with $r + s \leq 1$, then 
$
\mu_R(A\cup B) = \mu_R(A) + \mu(B) - \mu_R(A \cap B) \leq \mu_R(A) + \mu_R(B) < r + s,
$
which implies $\text{\ding{174}}$. 
\item If $\mu_R(A\cup B) \geq r$ and $\mu_R(A \cup (S\setminus B)) \geq s$, then 
$
\mu_R(A) = \mu_R(A\cup B) + \mu_R(A \cup (S\setminus B)) \geq r + s,
$
hence $\text{\ding{175}}$.
\item Assume $\mu_R(A) \geq r$ and $r + s > 1$, then 
$
\mu_R(S\setminus A) = \mu_R(S) - \mu_R(A) < p,
$
thus $\text{\ding{176}}$ holds. 
\item
If $\mu_R(\emptyset) \geq r$, then $r = 0$, yielding $\text{\ding{177}}$. 
\item 
Finally, if $\Folge{A}$ is decreasing with $\mu_R(A_n) \geq r$ for each $n\in\Nat$, then it is plain that
$
\mu_R(\bigcap_{n\in\Nat}A_n) \geq r.
$
This implies $\text{\ding{178}}$. 
\end{enumerate}
\EndProof

This permits a characterization of those stochastic effectivity functions which are generated through stochastic relations.

\BeginProposition{GtoK}
Let $P$ be a stochastic effectivity frame on state space $S$. Then these conditions are equivalent
\begin{enumerate}
\item\label{GtoK-1} There exists a stochastic relation $K: S \Trans S$ such that $P = P_K$.
\item\label{GtoK-2} $R(s) := \{\langle r, A\rangle \mid \bas{A}{\geq r}\in P(s)\}$ defines  a characteristic relation on $S$ with $P(s) \vdash R(s)$ for each state $s \in S$.
\end{enumerate}
\EndProposition

\BeginProof
``\labelImpl{GtoK-1}{GtoK-2}'':
Fix $s \in S$.  Because 
$
\bas{A}{\geq r}\in P_K(s)
$
iff
$
K(s)(A) \geq r,
$
we see that 
$
P(s) \models K(s),
$
hence by Proposition~\ref{satisf=implem}
$
P(s) \vdash R(s).
$

``\labelImpl{GtoK-2}{GtoK-1}'':
Define 
$
K(s) := \mu_{R(s)},
$
for $s \in S$, then $K(s)$ is a subprobability measure on (the Borel sets of) $S$. We show that $K: S \Trans S$. Let $G \in\Borel{\SubProb{S}}$ be a Borel set, then 
$
G \times [0, 1]\in\Borel{\SubProb{S}\otimes[0, 1]},
$
hence the measurability condition on $P$ yields that
\begin{equation*}
\InvBild{K}{G}  = \{s \in S \mid K(s)\in G\}
 = \{s \in S \mid G \in P(s)\}
\end{equation*}
is a measurable subset of $S$, because 
\begin{equation*}
\{\langle s, q\rangle \mid (G \times [0, 1])_q\in P(s)\} 
=
\{s \in S \mid G\in P(s)\} \times [0, 1]
\in \Borel{S\otimes[0, 1]}.
\end{equation*}
\EndProof

\subsection{Morphisms}
\label{sec:morphs}
Given stochastic effectivity functions $P$ on $S$ and $Q$ on $T$, a measurable map $f: S \to T$ is called a \emph{morphism of effectivity functions} $f: P \to Q$ iff this diagram commutes
\begin{equation*}
\xymatrix{
S\ar[d]_{P}\ar[rr]^f && T\ar[d]^{Q}\\
\Vau{S}\ar[rr]_{\Vau{f}} && \Vau{T}
}
\end{equation*}
Thus we have 
\begin{equation}
\label{morph-eff}
W \in Q(f(s))
\Leftrightarrow
\InvBild{\SubProb{f}}{W}\in P(s)
\end{equation}
for all states $s \in S$ and for all $W \in \Borel{\SubProb{T}}$.

Similarly, a measurable map $f: S \to T$ is a \emph{morphism of stochastic relations} $f: K \to L$ for the stochastic relations $K: S \Trans S$ and $L: T \Trans T$ iff this diagram commutes
\begin{equation*}
\xymatrix{
S\ar[d]_{K}\ar[rr]^f && T\ar[d]^{L}\\
\SubProb{S}\ar[rr]_{\SubProb{f}} && \SubProb{T}
}
\end{equation*}
Thus 
\begin{equation}
\label{morph-stoch}
L(f(s))(B) = \SubProb{f}(K(s))(B) = K(s)(\InvBild{f}{B})
\end{equation}
for each state $s\in S$ and each Borel set $B \in \Borel{T}$. 

These notions of morphisms are compatible: Each  morphism for stochastic relations turns into a morphism for the associated effectivity function (we will usually do without the attributions to effectivity functions or stochastic relations when talking about morphisms, whenever the context is clear). 
 
\BeginProposition{KripkeToGame}
A morphism $f: K \to L$ for stochastic relations $K$ and $L$ induces a morphism $f: P_K\to P_L$ for the associated stochastic effectivity functions.
\EndProposition

\BeginProof
Fix  a state $s\in S$. Then $W \in P_L(f(s))$ iff $L(f(s))\in W$. Because $f$ is a Kripke frame morphism, this is equivalent to $\SubProb{f}(K(s))\in W$, hence to $K(s)\in\InvBild{\SubProb{f}}{W}$, thus 
$
\InvBild{\SubProb{f}}{W} \in P_K(s).
$
\EndProof

\subsection{Congruences}
\label{sec:congr}

Morphisms and congruences are quite closely connected in algebraic systems, so after having defined and briefly investigated morphisms, we turn to congruences as those equivalence relations which are related to the structure of the underlying system. The most interesting equivalence relation for the purposes of the present paper is the one induced by the logic; it renders two states equivalent iff they satisfy exactly the same formulas. This will be investigated in Section~\ref{sec:induced}, the present section prepares for the discussion.

Let $P$ be a stochastic effectivity function on the $S$. Congruences are defined as usual through morphisms and factorization.  

\BeginDefinition{def-congr}
The equivalence relation $\rho$ on $S$ is called a \emph{congruence for $P$} iff there exists an effectivity function $P_\rho$ on $\Faktor{S}{\rho}$ which renders this diagram commutative
\begin{equation*}
\xymatrix{
S\ar[d]_P\ar[rr]^{\fMap{\rho}} && \Faktor{S}{\rho}\ar[d]^{P_\rho}\\
\Vau{S}\ar[rr]_{\Vau{\fMap{\rho}}}  && \Vau{\Faktor{S}{\rho}}
}
\end{equation*}
\EndDefinition

Because $\fMap{\rho}$ is onto, $P_\rho$ is uniquely determined. The next proposition provides a criterion for an equivalence relation to be a congruence. It requires the equivalence relation to be tame, so that quantitative aspects are being taken care of. 

\BeginProposition{is-a-congruence}
Let $\rho$ be a tame equivalence relation on $S$. Then these statements are equivalent
\begin{enumerate}
\item\label{is-a-congruence-1} $\rho$ is a congruence for $P$.
\item\label{is-a-congruence-2} Whenever $\isEquiv{s}{s'}{\rho}$, we have 
$
\InvBild{\SubProb{\fMap{\rho}}}{A} \in P(s) \text{ iff } \InvBild{\SubProb{\fMap{\rho}}}{A} \in P(s')
$
for every $A \in \bS{\Faktor{S}{\rho}}$ 
\end{enumerate}
\EndProposition

\BeginProof
``\labelImpl{is-a-congruence-1}{is-a-congruence-2}'': This follows immediately from the definition of a morphism, see~(\ref{morph-eff}). 

``\labelImpl{is-a-congruence-2}{is-a-congruence-1}'': Define for $s \in S$
\begin{equation*}
Q(\Klasse{s}{\rho}) := \{A \in \bS{\Faktor{S}{\rho}} \mid 
\InvBild{\SubProb{\fMap{\rho}}}{A} \in P(s)\},
\end{equation*}
then $Q$ is well defined by the assumption, and it is clear that $Q(\Klasse{s}{\rho})$ is an upward closed set of subsets of  $\bS{\Faktor{S}{\rho}}$ for each $s \in S$. It remains to be shown that $Q$ is a stochastic effectivity function, i.e., that $Q$ is t-measurable. In fact, let 
$
H \in \Borel{\SubProb{\Faktor{S}{\rho}}\otimes[0, 1]}
$
be a test set, and let 
$
G := \InvBild{(\SubProb{\fMap{\rho}}\times id_{[0, 1]})}{H}
$
be its inverse image under $\SubProb{\fMap{\rho}}\times id_{[0, 1]}$, then
\begin{equation*}
\{\langle t, q\rangle \in \Faktor{S}{\rho}\times[0, 1] \mid H_q \in Q(t)\}
= 
\Bild{(\fMap{\rho}\times id_{[0, 1]})}{Z}.
\end{equation*}
with 
$
Z := \{\langle s, q\rangle \in S\times[0, 1] \mid G_q \in P(s)\}.
$
By Corollary~\ref{cor-is-tame} it is enough to show that $Z$ is contained in $\Inv{\rho}{S}\otimes\Borel{[0, 1]}$. Because $P$ is t-measurable, we infer 
$
Z \in \Borel{S\otimes[0, 1]},
$
and because $Z$ is $(\rho\times\Delta)$-invariant, we conclude that 
$
Z \in \Inv{\rho\times\Delta}{S\otimes[0, 1]},
$
the latter $\sigma$-algebra being  equal to 
$
\Inv{\rho}{S}\otimes\Borel{[0, 1]}
$
by Lemma~\ref{is-tame}. 
\EndProof

The condition on subsets of $\bS{\Faktor{S}{\rho}}$ imposed above asks
for Borel sets of $\SubProb{\Faktor{S}{\rho}}$, so factorization is
done ``behind the curtain'' of functor $\SubProbSenza$. It would be more
convenient if the space of all subprobabilities could be factored and
the corresponding Borel sets formed on the latter space. In fact, lift
equivalence $\rho$ on $S$ to an equivalence $\bar{\rho}$ on
$\SubProb{S}$ upon setting
\begin{equation*}
\isEquiv{\mu}{\mu'}{\bar{\rho}}\text{ iff } \forall C \in \Inv{\rho}{S}: \mu(C) = \mu'(C),
\end{equation*}
so measures are considered $\bar{\rho}$-equivalent iff they coincide on the $\rho$-invariant measurable sets. Define
the map $\partial_\rho$ through
\begin{equation*}
\begin{cases}
\Faktor{\SubProb{S}}{\bar{\rho}} &\to \SubProb{\Faktor{S}{\rho}}\\
\partial_\rho(\Klasse{\mu}{\bar{\rho}})&\mapsto \lambda G.\SubProb{\fMap{\rho}}(\mu)(G)
\end{cases}
\end{equation*} 
Then 
$
\partial_\rho\circ\fMap{\bar{\rho}} = \SubProb{\fMap{\rho}}, $ so that
$\partial_\rho$ is measurable by finality of $\fMap{\bar{\rho}}$. If
$S$ is a Polish space, and $\rho$ is countably generated (thus
$\Inv{\rho}{S} = \sigma(\{A_n \mid n \in \Nat\})$ for some sequence
$\Folge{A}$ with measurable $A_n$), then it can be shown through
Souslin's Separation Theorem that $\partial_\rho$ is an
isomorphism~\cite[Section 1.8.1]{EED-CoalgLogic-Book}. Hence it is in
this case sufficient to focus on the sets $
\InvBild{\fMap{\bar{\rho}}}{W} $ with $W \in
\Borel{\Faktor{\SubProb{S}}{\bar{\rho}}}$.
The structural question of characterizing the subprobabilities of a factor space through invariant sets will be taken up again in Lemma~\ref{borel-isom}.

The relationship of morphisms and congruences through the kernel of the morphism is characterized now. It assumes the morphism combined with the identity on $[0, 1]$ to be final. This is a technical condition rendering the kernel of the morphism a tame equivalence relation. 

\BeginProposition{final-is-congr}
Given a morphism $f: P \to Q$ for the effectivity functions $P$ and $Q$ over the state spaces $S$ resp. $T$. If $f\times id_{[0, 1]}: S\times [0, 1] \to T\times [0, 1]$ is final, then $\Kern{f}$ is a congruence for $P$.  
\EndProposition
{
\def\fKf{\fMap{\Kern{f}}}
\def\xKf{\Faktor{S}{\Kern{f}}}
\BeginProof
0.
We show first that the equivalence relation $\Kern{f\times id_{[0, 1]}} = \Kern{f}\times\Delta$ is tame. 
It is easy to see that $f: S \to T$ is final (because $D\times[0, 1]\in\Borel{T\otimes[0, 1]}$ iff $D\in\Borel{T}$), and we infer from Lemma~\ref{kerf-is-exact} that
\begin{equation*}
\InvBild{(f\times id_{[0, 1]})}{\Borel{S \otimes [0, 1]}} = \Inv{\Kern{f}\times\Delta}{S\otimes[0, 1]},
\end{equation*}
on the other hand, 
\begin{equation*}
\InvBild{(f\times id_{[0, 1]})}{\Borel{S \otimes [0, 1]}} 
\subseteq 
\InvBild{f}{\Borel{T}}\otimes\Borel{[0, 1]}
=
\Inv{\Kern{f}}{S}\otimes\Borel{[0, 1]}.
\end{equation*}
Thus 
$
\Inv{\Kern{f}\times\Delta}{S\otimes[0, 1]} = \Inv{\Kern{f}}{S}\otimes\Borel{[0, 1]},
$
hence $\Kern{f}\times\Delta$ is tame. 

1.
Given $H_0\in\bS{\xKf}$, we claim that we can find $H\in\bS{T}$ such that 
$
H_0 = \InvBild{\SubProb{\widetilde{f}}}{H},
$
$f = \widetilde{f}\circ\fMap{\Kern{f}}$ being the decomposition of $f$ according to the proof of Lemma~\ref{kerf-is-exact}. 
In fact, put
\begin{equation*}
\mathcal{Z} := \{H_0\in\bS{\xKf} \mid \exists H\in\bS{T}: H_0 = \InvBild{\SubProb{\widetilde{f}}}{H}\}.
\end{equation*}
Then $\mathcal{Z}$ is a $\sigma$-algebra, because $\emptyset\in\mathcal{Z}$ and it is closed under the countable Boolean operations as well as complementation. Let 
$
\basS{\xKf}{A}{\geq q}
$ 
with $A \in \Borel{\xKf}$ be an element of the basis for the weak-*-$\sigma$-algebra. Because $\widetilde{f}^{-1}$ is onto, there exists $B \subseteq T$ with 
$
A = \Bild{\widetilde{f}}{B},
$
by the remark above we know that $B \in \Borel{T}$. Consequently, 
$
\mu(A) = \SubProb{\widetilde{f}}(\mu)(B)
$
for any 
$
\mu \in \SubProb{\xKf}.
$
This implies
$
\basS{\xKf}{A}{\geq q} = \InvBild{\SubProb{\widetilde{f}}}{\basS{T}{B}{\geq q}} \in \mathcal{Z},
$
consequently, 
\begin{equation*}
\bS{\xKf} = \sigma(\{\basS{\xKf}{A}{\geq q} \mid A \in \Borel{\xKf}, q \geq 0\}) \subseteq \mathcal{Z},
\end{equation*}
thus
$
\bS{\xKf} = \mathcal{Z}.
$

2.
Now let $f(s) = f(s')$, and take $H_0\in\bS{\xKf}$, choose $H\in\bS{T}$ according to part 1. for $H_0$, then 
\begin{align*}
\InvBild{\fKf}{H_0} \in P(s)
& \Leftrightarrow
\InvBild{f}{H} \in P(s) \\
& \Leftrightarrow
H \in Q(f(s)) = Q(f(s'))\\
& \Leftrightarrow
\InvBild{f}{H} \in P(s') \\
& \Leftrightarrow
\InvBild{\fKf}{H_0} \in P(s'),
\end{align*}
because $f:P  \to Q$ is a morphism.
\EndProof
}

Let us turn to the case of stochastic relations and investigate briefly the relationship of congruences and the effectivity functions generated through the factor relation. A \emph{congruence $\varpi$ for a stochastic relation} $K: S \Trans S$ is an equivalence relation with this property: There exists a (unique) stochastic relation $K_\varpi: \Faktor{S}{\varpi}\to\Faktor{S}{\varpi}$ such that this diagram commutes
\begin{equation*}
\xymatrix{
S\ar[rr]^{\fMap{\varpi}}\ar[d]_K && \Faktor{S}{\varpi}\ar[d]^{K_\varpi}\\
\SubProb{S}\ar[rr]_{\SubProb{\fMap{\varpi}}} && \SubProb{\Faktor{S}{\varpi}}
}
\end{equation*}
This translates to
\begin{equation*}
K_\varpi(\Klasse{s}{\varpi})(B) = K(s)(\InvBild{\fMap{\varpi}}{B})
\end{equation*}
for all $s \in S$ and all $B \in \Borel{\Faktor{S}{\varpi}}$, see~\cite[Section 1.7.3]{EED-CoalgLogic-Book}.  We obtain from Proposition~\ref{KripkeToGame}:
\BeginCorollary{stochCongr-is-effCongr}
A congruence $\varpi$ for a stochastic relation $K: S \Trans S$ is also a congruence for the associated effectivity function $P_K$. Moreover, 
$
P_{K_\varpi} = (P_K)_\varpi,
$
so the effectivity function associated with the factor relation $K_\varpi$ is the factor relation of $P_K$ with respect to $\varpi$.
\QED
\EndCorollary

It is noted that we do not require additional assumptions on the congruence for the stochastic relation for being a congruence for the associated effectivity function. This indicates that the condition on tameness captures the general class of effectivity functions, but that subclasses may impose their own conditions. It indicates also that the condition of being a congruence for a stochastic relation itself is a fairly strong one when assessed by the rules pertaining to stochastic effectivity functions.

%
%


\section{Games Frames and Transformations}
\label{sec:gameFr-Trans}
A game frame has a state space and assigns to each primitive game a stochastic effectivity function. Game frames will later be extended to models for game logic, for the time being, however, we focus on what can be said about composite games, when viewed through the glasses of a game frame. We will define game frames and assign through a game frame to each game $\tau$ the set  $A$ of states such that Angel has a strategy of reaching $A$ by playing $\tau$. This yields a family of set transformers for each game $\tau$. We show that the transformed states are always measurable, provided the state space is closed under Souslin's operation~(\ref{Souslin}). The special case that the game frame is generated by a stochastic Kripke model is discussed as well, and it is shown that the basic probabilities can be computed through the convolution of the corresponding relations. We have also a look at morphisms and consider the relationship of morphisms with the sets of states computed for  games.

\BeginDefinition{G-frame}
A \emph{game frame} $\Ge = (S, (P_\gamma)_{\gamma\in\Gamma})$ has a measurable space $S$ of states and  a t-measurable map $P_\gamma: S \to \Vau{S}$ for each primitive game $\gamma\in\Gamma$.
\EndDefinition

Fix a game frame 
$
\Ge = (S, (P_\gamma)_{\gamma\in\Gamma}),
$
the set of primitive games is extended by the empty game $\epsilon$, and set 
$
P_\epsilon := D
$
with $D$ as the Dirac effectivity function according to Example~\ref{dirac}. We assume in the sequel that $\epsilon\in\Gamma$.

\medskip

When writing down games, we assume for simplicity that composition binds tighter than angelic or demonic choice. We make these assumptions~\cite{Pauly-Parikh,Parikh-Games1985}:
\begin{enumerate}[i.]
\item $(\tau^d)^d$ is identical to $\tau$.
\item Demonic choice can be represented through angelic choice: The game $\tau_1\cap\tau_2$ coincides with the game $(\tau_1^d\cup\tau_2^d)^d$.
\item Similarly, demonic iteration can be represented through its angelic counterpart: $(\tau^\times)^d$ is equal to $(\tau^d)^*$,
\item Composition is right distributive with respect to angelic choice: Making a decision to play $\tau_1$ or $\tau_2$ and then playing $\tau$ should be the same as deciding to play $\tau_1;\tau$ or $\tau_2;\tau$, thus $(\tau_1\cup\tau_2);\tau$ equals $\tau_1;\tau\cup\tau_2;\tau$. 

Note that left distributivity would mean that a choice between $\tau;\tau_1$ and $\tau;\tau_2$ is the same as playing first $\tau$ then $\tau_1\cup\tau_2$; this is a somewhat restrictive assumption, since the choice of playing $\tau_1$ or $\tau_2$ may be a decision made by Angel only after $\tau$ is completed~\cite[p. 191]{Benthem-LogicGames}. Thus we do not assume this in general (it will turn out, however, that in Kripke generated models these choices are in fact equivalent, see Section~\ref{sec:distr-prog}).
\item\label{star-assumption}
We assume similarly that 
$
\tau^*;\tau_0 
\text{ equals }
\tau;\tau_0\cup\tau^*;\tau;\tau_0.
$
Hence when playing $\tau^*;\tau_0$  Angel may decide to play $\tau$ not at all and to continue with $\tau_0$ right away, or to play $\tau^*$ followed by $\tau;\tau_0$. Thus $\tau^*;\tau_0$ expands to 
$
\tau_0 \cup \tau;\tau_0 \cup \tau;\tau;\tau_0 \cup \dots.
$
\item $(\tau_1;\tau_2)^d$ is the same as $\tau_1^d;\tau_2^d$.
\item The binary operations (composition, angelic and demonic choice) are commutative and associative. 
\end{enumerate}

We define now recursively the set valued function $\dasI{\tau}{A}{q}$ with the informal meaning that this set describes the set of states so that Angel has a strategy of reaching a state in set $A$ with probability greater than $q$ upon playing game $\tau$. Assume that $A\in\Borel{S}$ is a measurable subset of $S$, and $0 \leq q < 1$, and define for $0 \leq k \leq \infty$
\begin{equation*}
Q^{(k)}(q) := \{\langle a_1, \dots, a_k\rangle \in \Rational^k \mid a_i \geq 0\text{ and } \sum_{i=1}^k a_i \leq q\}.
\end{equation*}
as the set of all non-negative rational $k$-tuples the sum of which does not exceed $q$.
\begin{enumerate}[A.]
\item\label{primitive} Let $\gamma\in\Gamma$, then put
\begin{equation*}
\dasI{\gamma}{A}{q} := \{s\in S \mid \bas{A}{> q}\in P_\gamma(s)\}, 
\end{equation*}
in particular $\dasI{\epsilon}{A}{q} = \{s \in S \mid \delta_s(A) > q\} = A$. Thus $s \in \dasI{\gamma}{A}{q}$ iff Angel has $\bas{A}{> q}$ in its portfolio when playing $\gamma$ in state $s$. This entails that the set of all state distributions which evaluate at $A$ with a probability greater than $q$ can be effected by Angel in this situation. If Angel does not play at all, hence if the game $\gamma$ equals $\epsilon$, nothing is about to change, which means 
$
\dasI{\epsilon}{A}{q} = \{s \mid \delta_s\in\bas{A}{> q}\} = A.
$
\item\label{dual} Let $\tau$ be a game, then 
\begin{equation*}
\dasI{\tau^d}{A}{q} := S \setminus\dasI{\tau}{S \setminus A}{q}.
\end{equation*}
The game is determined, thus Demon can reach a set of states iff Angel does not have a strategy for reaching the complement. Consequently, upon playing $\tau$ in state $s$, Demon can reach a state in $A$ with probability greater than $q$ iff Angel cannot reach a state in $S\setminus A$ with probability greater $q$. 

Illustrating, let us assume for the moment that $P_\gamma = P_{K_\gamma}$, i.e., that the effectivity function for $\gamma\in\Gamma$ is generated from a stochastic relation $K_\gamma$. Then 
\begin{equation*}
s \in \dasI{\gamma^d}{A}{q}
\Leftrightarrow
s \notin \dasI{\gamma}{S \setminus A}{q} 
\Leftrightarrow
K_\gamma(s)(S\setminus A) \leq q. 
\end{equation*}
In general, 
$
s \in \dasI{\gamma^d}{A}{q}
$
iff 
$
\bas{S\setminus A}{> q} \notin P_\gamma(s)
$
for $\gamma\in\Gamma$. This is exactly what one would expect in a determined game.
\item\label{angelic-choice}
Assume $s$ is a state such that Angel has a strategy for reaching a state in $A$ when playing the game $\tau_1\cup\tau_2$ with probability not greater than $q$. Then  Angel should have a strategy in $s$ for reaching a state in $A$ when playing game  $\tau_1$ with probability not greater than $a_1$ and playing game  $\tau_2$ with probability not greater than $a_2$ such that $a_1 + a_2 \leq q$. Thus
\begin{equation*}
\dasI{\tau_1\cup\tau_2}{A}{q} := \bigcap_{a \in Q^{(2)}(q)}\bigl(\dasI{A}{\tau_1}{a_1}\cup\dasI{A}{\tau_2}{a_2}\bigr).
\end{equation*}
\item\label{Left-distr} 
Right distributivity of composition over angelic choice translates to this equation.
\begin{equation*}
\dasI{(\tau_1\cup\tau_2);\tau}{A}{q} := \dasI{\tau_1;\tau\cup\tau_2;\tau}{A}{q}. 
\end{equation*}
\item\label{Left-times} If $\gamma\in\Gamma$, put
\begin{equation*}
\dasI{\gamma;\tau}{A}{q} := \{s \in S \mid G_\tau(A, q) \in P_\gamma(s)\},
\end{equation*}
where
\begin{equation}
\label{def_g}
G_\tau(A, q) := \{\mu\in\SubProb{S} \mid \int_0^1\mu(\dasI{\tau}{A}{r})\ dr > q\}.
\end{equation}
Suppose that $\dasI{\tau}{A}{r}$ is already defined for each $r$ as the set of states for which Angel has a strategy to effect a state in $A$ through playing $\tau$ with probability greater than $r$. Given a distribution $\mu$ over the states, the integral 
$
\int_0^1\mu(\dasI{\tau}{A}{r})\ dr
$
is the expected value for entering a state in $A$ through playing $\tau$ for $\mu$. The set $G_\tau(A, q)$ collects all distributions the expected value of which is greater that $q$. We ask for all states such that Angel has this set in its portfolio when playing $\gamma$ in this state. Being able to select this set from the portfolio means that when playing $\gamma$ and subsequently $\tau$ a state in $A$ may be reached with probability greater than $q$.
\item\label{star}
This is just the translation of \ref{star-assumption}. with a repeated application of the rule \ref{angelic-choice}. for angelic choice:
\begin{equation*}
\dasI{\tau^*;\tau_0}{A}{q} := \bigcap_{a\in Q^{(\infty)}(q)}\bigcup_{n\geq 0}\dasI{\tau^n;\tau_0}{A}{a_{n+1}}
\end{equation*}
with $\tau^n := \tau;\dots;\tau$ ($n$ times).
\end{enumerate}

It has to be established that $\dasI{\tau}{A}{q}\in\Borel{S}$, provided $A\in\Borel{S}$. We look at different cases.

\BeginLemma{measurb-1}
Let $\tau$ be a game such that
$
\{\langle s, r\rangle \in S\times [0, 1] \mid s \in \dasI{\tau}{A}{r}\}
$ 
is a measurable subset of $S\times [0, 1]$, and assume that $\gamma\in\Gamma$. Then  
$
\{\langle s, r\rangle \in S\times [0, 1] \mid s \in\dasI{\gamma; \tau}{A}{r}\}
$
is a measurable subset of $S\times [0, 1]$. 
\EndLemma

\BeginProof
This follows from Corollary~\ref{meas-3}, because  $P_\gamma$ is t-measurable.
\EndProof

The transformation associated with the indefinite iteration in~\ref{star} above involves an uncountable intersection, since for $q > 0$ the set $Q^{(\infty)}(q)$ has the cardinality of the continuum. Since $\sigma$-algebras are closed only under countable operations, we might generate in this way a set which is not measurable at all, provided we do not take cautionary measures. If the state space is closed under the \emph{Souslin operation}, see~\ref{Souslin} in the Section~\ref{sec:indispensable}, it can be shown that the resulting set will still be measurable. A fairly popular class of spaces closed under this operation is the class of universally complete measurable spaces, see Section~\ref{sec:complete}. 

\BeginLemma{measurb-2}
Let $\tau$ and $\tau_0$ be games such that 
$
\dasI{\tau^n;\tau_0}{A}{r}
$
is a measurable subset of $S$ for each $n\in\Nat$ and each $r \in [0, 1]$. Assume that $S$ is closed under the Souslin operation. Then 
$
\dasI{\tau^*;\tau_0}{A}{q}
$
is a measurable subset of $S$ for each $n\in\Nat$ and each $r \in [0, 1]$.
\EndLemma

\BeginProof
This follows as in~\cite[Proposition 6.7]{EED-PDL-TR}.
\EndProof

Call the game $\tau$ \emph{interpretable} iff $\dasI{\tau}{A}{q}$ is defined for each $A\in\Borel{S}, q \in[0, 1]$ so that the set 
\begin{equation*}
Gr(\tau, A) := \{\langle s, q\rangle\in S \times [0, 1] \mid s \in \dasI{\tau}{A}{q}\}
\end{equation*}
 is a measurable subset of $S \times [0, 1]$. 

\BeginLemma{interpretable}
Each game $\tau$ is interpretable, provided the state space is closed under the Souslin operation.
\EndLemma

\BeginProof
1.
Let 
$
J := \{\tau \mid \tau\text{ is interpretable}\}.
$
We show that $J$ contains all games.

2.
Let $A\in\Borel{S}$, then
\begin{equation*}
\{\langle \mu, q\rangle\in \SubProb{S} \times [0, 1] \mid \mu\in\bas{A}{> q}\}
= 
\{\langle \mu, q\rangle\in \SubProb{S} \times [0, 1] \mid \mu(A) > q\}
\end{equation*}
is a measurable subset of $\SubProb{S}\times[0, 1]$, see Corollary~\ref{meas-2}. Thus, if $\gamma \in \Gamma$, we have 
\begin{equation*}
Gr(\gamma, A) = \{\langle s, q\rangle\in S \times [0, 1] \mid \bas{A}{> q} \in P_\gamma(s)\},
\end{equation*}
which is a measurable subset of $S \times [0, 1]$. Consequently, $\Gamma\subseteq J$. 

3.
Clearly, $J$ is closed under demonization and angelic choice, hence under demonic choice as well. Now let 
\begin{equation*}
L := \{\tau \mid \tau;\tau_1\text{ is interpretable for all interpretable }\tau_1\}.
\end{equation*}
Then Lemma~\ref{measurb-1} implies that $\Gamma\cup\{\gamma^d\mid\gamma\in\Gamma\}\subseteq L$. Moreover, because angelic choice distributes from the left over composition, $L$ is closed under angelic choice. It is also closed under demonization: Let $\tau\in L$ and take an interpretable game $\tau_1$, then $\tau;\tau_1^d$ is interpretable, thus the interpretation of $(\tau;\tau_1^d)^d$ is defined, hence $\tau^d; \tau_1$ is interpretable. Clearly, $L$ is closed under composition. Thus $\tau\in L$ implies $\tau^*\in L$ as well as $\tau^\times\in L$, so that $L$ is the set of all games. 

This implies that $J$ is closed under composition, and hence both under angelic and demonic iteration.     
\EndProof

This yields
\BeginProposition{prop-interp}
Assume that the state space is closed under the Souslin operation, then we have 
$ 
\dasI{\tau}{A}{q} \in \Borel{S}
$ 
for all games $\tau$, $A\in\Borel{S}$ and $0 \leq q \leq 1$. 
\EndProposition

\BeginProof
We infer for each game $\tau$ from Lemma~\ref{interpretable} that for $A\in\Borel{S}$ the set $Gr(\tau, A)$ is a measurable subset of $S \times [0, 1]$. But 
$\dasI{\tau}{A}{q} = Gr(\tau, A)_q$.
\EndProof

\medskip

Suppose that $\mathcal{H} = (T, (Q_\gamma)_{\gamma\in \Gamma})$ is another game frame, then $f: \mathcal{G} \to \mathcal{H}$ is a \emph{game frame morphism } iff $f: P_\gamma\to Q_\gamma$ is a morphism for the associated effectivity functions for all $\gamma\in\Gamma$. The transformations above are compatible with frame morphisms.

\BeginProposition{g-morph-compatible}
Let $f: \mathcal{G} \to \mathcal{H}$ be a game frame morphism, and assume that $\dasI[\mathcal{G}]{\tau}{\cdot}{q}$ and $\dasI[\mathcal{H}]{\tau}{\cdot}{q}$ always transforms measurable sets into measurable sets for all games $\tau$ and all $q$. Then we have 
\begin{equation*}
\InvBild{f}{\dasI[\mathcal{H}]{\tau}{B}{q}} = \dasI[\mathcal{G}]{\tau}{\InvBild{f}{B}}{q}
\end{equation*} 
for all games $\tau$, all measurable sets $B\in\Borel{T}$ and all $q$.
\EndProposition

\BeginProof
0.
The proof proceeds by induction on $\tau$. Because $f$ is a morphism, the assertion is true for $\tau = \gamma\in\Gamma$. Because $f^{-1}$ is compatible with the Boolean operations on sets, it is sufficient to consider the case $\tau = \gamma;\tau_1$ in detail. 

1.
Assume that the assertion is true for game $\tau_1$, fix $B\in\Borel{T}, q \geq 0$. Then 
\begin{align*}
G_{\tau_1, \mathcal{G}}(\InvBild{f}{B}, q) 
:= &\ \{\mu\in\SubProb{S} \mid \int_0^1\mu\bigl(\dasI{\tau_1}{\InvBild{f}{B}}{r})\ dr > q\}\\
 \stackrel{(\star)}{=} &\ \{\mu\in\SubProb{S} \mid \int_0^1\mu\bigl(\InvBild{f}{\dasI{\tau_1}{B}{r}}\bigr)\ dr > q\}\\
 \stackrel{(\oplus)}{=} &\ \{\mu\in\SubProb{S} \mid \int_0^1\SubProb{f}(\mu)\bigl(\dasI[\mathcal{H}]{\tau_1}{B}{r}\bigr)\ dr > q\}\\
 = &\ \InvBild{\SubProb{f}}{\{\nu\in\SubProb{T} \mid \int_0^1\nu\bigl(\dasI[\mathcal{H}]{\tau_1}{\InvBild{f}{B}}{r}\bigr)\ dr > q\}}\\
 = &\  \InvBild{\SubProb{f}}{G_{\tau_1, \mathcal{H}}(B, q)}
\end{align*}
The equation $(\star)$ derives from the induction hypothesis, and $(\oplus)$ from the definition of $\SubProb{f}(\mu)$. 

2.
Because $f: P_\gamma\to Q_\gamma$ is a morphism, we obtain now
\begin{align*}
\dasI{\gamma;\tau_1}{\InvBild{f}{B}}{q}
= &\ \{s \in S \mid G_{\tau_1, \mathcal{G}}(\InvBild{f}{B}, q) \in P_\gamma(s)\}\\
= &\ \{s \in S \mid \InvBild{\SubProb{f}}{G_{\tau_1, \mathcal{H}}(B, q)} \in P_\gamma(s)\}\\
= &\ \{s \in S \mid G_{\tau_1, \mathcal{H}}(B, q)\in Q_\gamma(f(s))\}\\
= &\ \InvBild{f}{\dasI[\mathcal{H}]{\gamma;\tau_1}{B}{q}}
\end{align*}
This shows that the assertion is also true for $\tau = \gamma; \tau_1$. 
\EndProof

Let us briefly interpret Proposition~\ref{g-morph-compatible} in terms of natural transformations. 

\BeginExample{nat-transf}
Fix a game $\tau$ and a real $q \in [0, 1]$, then $\dasI{\tau}{\cdot}{q}: \Borel{S}\to\Borel{S}$ by Proposition~\ref{prop-interp}, provided $S$ satisfies the Souslin condition. Let us briefly assume that $\BorelSenza$ acts as a contravariant functor from the category of measurable spaces satisfying the Souslin condition to the category of sets, where the measurable map $f: S \to T$ is mapped to $\Borel{f}: \Borel{T}\to\Borel{S}$ by $\Borel{f} := f^{-1}$. Then $\dasI{\tau}{\cdot}{q}$ induces a natural transformation $\BorelSenza\to\BorelSenza$, because by Proposition~\ref{g-morph-compatible} this diagram commutes:
\begin{equation*}
\xymatrix{
S\ar[d]_{f}&&\Borel{T}\ar[d]_{\dasI[\mathcal{H}]{\tau}{\cdot}{q}}\ar[rr]^{f^{-1}}&&\Borel{S}\ar[d]^{\dasI{\tau}{\cdot}{q}}\\
T          &&\Borel{T}\ar[rr]_{f^{-1}}                                           &&\Borel{S} 
}
\end{equation*}
\EndExample

We turn now to the special case of Kripke generated frames.

\subsection{Kripke generated frames}

A \emph{stochastic Kripke frame} $\Ke = (S, (K_\gamma)_{\gamma\in\Gamma})$ is a measurable state space $S$, each primitive game $\gamma\in\Gamma$ is associated with a stochastic relation  $K_\gamma: S \Trans S$. Morphisms carry over in the obvious fashion from stochastic relations to stochastic Kripke frames by applying the defining condition to the stochastic relation associated with each primitive game. 

We associate with $\Ke$ a game frame $\Ge_\Ke := (S, (P_{K_\gamma})_{\gamma\in\Gamma})$. Thus the transformations associated with games considered above are also applicable to Kripke models. We will discuss this shortly. An application of Proposition~\ref{GtoK} for each $\gamma\in\Gamma$ states under which conditions a game frame is generated by a stochastic Kripke frame. Just for the record:

\BeginProposition{GtoK-mod}
Let $\Ge = (S, (P_\gamma)_{\gamma\in\Gamma})$ be a game frame. Then these conditions are equivalent
\begin{enumerate}
\item\label{GtoK-1-mod} There exists a stochastic game frame $\Ke$ with $\Ge = \Ge_\Ke$.
\item\label{GtoK-2-mod} $R_\gamma(s) := \{\langle r, A\rangle \mid \bas{A}{\geq r}\in P_\gamma(s)\}$ defines  a characteristic relation on $S$ such that $P_\gamma(s) \vdash R_\gamma(s)$ for each state $s \in S, \gamma\in\Gamma$.
\QED
\end{enumerate}
\EndProposition

Let $\Ke = (S, (K_\gamma)_{\gamma\in\Gamma})$ be a Kripke frame with associated game frame $\Ge_{\Ke}$. $K_\gamma: S \Trans S$ are Kleisli morphisms; we  define their product ---~sometimes called convolution~--- through
\begin{equation*}
(K_{\gamma_1}\star K_{\gamma_2})(s)(A) := \int_S K_{\gamma_2}(t)(A)\ K_{\gamma_1}(s)(dt),
\end{equation*} 
see~\cite{Giry,EED-CoalgLogic-Book}. Intuitively, this gives the probability of reaching a state in $A\in\Borel{S}$, provided we start with game $\gamma_1$ in state $s$ and continue  with game $\gamma_2$, averaging over intermediate states (here $\gamma_1, \gamma_2\in\Gamma$). The observation that composing stochastic relations models the composition of modalities is one of the cornerstones for the interpretation of modal logics through Kripke models~\cite{Panangaden-book,EED-CoalgLogic-Book}. 

Let $\dasI{A}{\tau}{q}$ be defined as above when working in the game frame associated with Kripke frame $\Ke$. It turns out that  $\dasI{A}{\gamma_1; \dots \gamma_k}{q}$ can be described in terms of the Kleisli product for $K_{\gamma_1}, \dots, K_{\gamma_k}$, provided $\gamma_1, \dots, \gamma_k\in\Gamma$ are primitive games. 

\BeginProposition{primitive-product}
Assume that $\gamma_1, \dots, \gamma_k \in \Gamma$, then this equality holds in the game frame $\mathcal{G}$ associated with the Kripke frame $\Ke$
\begin{equation*}
\dasI{A}{\gamma_1; \dots \gamma_k}{q} = \{s\in S \mid (K_{\gamma_1}\star\dots\star K_{\gamma_k})(s)(A) > q\}.
\end{equation*}
for all $A\in\Borel{S}, 0 \leq q < 1$.
\EndProposition

\BeginProof
1.
The proof proceeds by induction on $k$. If $k = 1$, we have
\begin{equation*}
s \in \dasI{A}{\gamma_1}{q}
\Leftrightarrow
\bas{A}{> q} \in P_{\Ke, \gamma_1}(s)
\Leftrightarrow
K_{\gamma_1}(s)(A) > q.
\end{equation*}

2.
Assume that the claim is established for $k$, and let $\gamma_0\in\Gamma$. Then, borrowing the notation from above, 
\begin{align*}
s \in \dasI{A}{\gamma_0; \gamma_1; \dots; \gamma_k}{q}
& \Leftrightarrow
G_{\gamma_1; \dots; \gamma_k}(A, q) \in P_{\Ke, \gamma_0}(s)\\
& \Leftrightarrow
K_{\gamma_0}(s)\in G_{\gamma_1; \dots; \gamma_k}(A, q)\\
& \stackrel{}{\Leftrightarrow}
\int_0^1 K_{\gamma_0}(s)(\dasI{A}{\gamma_1; \dots; \gamma_k}{r})\ dr > q\\
& \stackrel{(\dagger)}{\Leftrightarrow}
\int_0^1 K_{\gamma_0}(s)(\{t \mid (K_{\gamma_1}\star\dots\star K_{\gamma_k})(t)(A) > r\})\ dr > q\\ 
& \stackrel{(\ddag)}{\Leftrightarrow}
\int_S (K_{\gamma_1}\star\dots\star K_{\gamma_k})(t)(A)\ K_{\gamma_0}(s)(dt) > q\\
& \stackrel{(\|)}{\Leftrightarrow}
(K_{\gamma_0}\star K_{\gamma_1}\star\dots\star K_{\gamma_k})(s)(A) > q
\end{align*}
Here $(\dagger)$ marks the induction hypothesis, $(\ddag)$ is the application of Choquet's Theorem~\ref{Choquet}, and $(\|)$ is the definition of the Kleisli product. This establishes the claim for $k+1$. 
\EndProof

We note as a consequence that the respective definitions of state transformations through the games under consideration coincide for game frames generated by Kripke frames. On the other hand it is noted that the definition of these transformations for general frames extends the one which has been used for Kripke frames for general modal logics.

\subsection{Distributivity in the PDL fragment}
\label{sec:distr-prog}

The games which are described through grammar~(\ref{prog-grammar}) with ($\Pi = \Gamma$) are called the \emph{PDL fragment}, the corresponding games are called \emph{programs} for simplicity. We will show now that in this fragment
\begin{equation*}
\dasI{\cdot}{\tau_1;(\tau_2\cup\tau_3)}{\cdot} = \dasI{\cdot}{\tau_1;\tau_2\cup\tau_1;\tau_3)}{\cdot}
\end{equation*}
holds, provided  frame $\Ge$ is generated by a stochastic Kripke frame $\Ke$. 

Define $\Meas{S}$ as the set of non-negative measures on (the measurable sets of) $S$ to the extended non-negative reals $\iReal := \pReal\cup\{\infty\}$ for the measurable space $S$. $\Meas{S}$ is closed under addition and under multiplication with non-negative reals; it is also closed under countable sums: given $\Folge{\mu}$ with $\mu_n\in\Meas{S}$, put
\begin{equation*}
\bigl(\sum_{n\in\Nat} \mu_n\bigr)(A) := \sup_{n\in\Nat} \sum_{i \leq n} \mu_i(A).
\end{equation*} 
Then $\sum_{n\in\Nat} \mu_n$ is monotone and $\sigma$-additive with $\bigl(\sum_{n\in\Nat} \mu_n\bigr)(\emptyset) = 0$, hence a member of $\Meas{S}$.

Call a map $N: S \to\Meas{S}$ an \emph{extended kernel} iff for each $A \in \Borel{S}$ the map $s \mapsto N(s)(A)$ is measurable (a subset $G\subseteq\iReal$ is called measurable iff $G\cap\pReal$ is a Borel set in $\pReal$). Extended kernels are closed under convolution: Put 
\begin{equation*}
(N_1\star N_2)(s)(A) := \int_S N_2(t)(A)\ N_1(s)(dt),
\end{equation*}
then $N_1\star N_2$ is an extended kernel again. This is  but the Kleisli composition applied to extended kernels. Thus $\Meas{S}$ is closed under convolution which distributes both from the left and from the right under addition and under scalar multiplication. Note that the countable sum of extended kernels is an extended kernel as well. 

Define recursively for the stochastic relations in the Kripke frame $\Ke$
\begin{align*}
K_{\tau_1\cup\tau_2} & := K_{\tau_1} + K_{\tau_2},\\
K_{\tau_1;\tau_2} & := K_{\tau_1} \star K_{\tau_2},\\
K_{\tau^*} & := \sum_{n\geq 0} K_{\tau^n}.
\end{align*} 
This defines $K_\tau$ for each $\tau$ in the PDL-fragment of game logic~\cite{Kozen-ProbPDL}.

Define 
\begin{equation*}
\dasL{A}{\tau}{q} := S \setminus \dasI{A}{\tau}{q},
\end{equation*}
where $\Ge$ is the game frame associated with the Kripke frame $\Ke$ over state space $S$, $A \in \Borel{S}$ is a measurable set, $\tau$ is a program, i.e., a member of the PDL fragment, and $q \in [0, 1]$. It is more convenient to work with these complements, as we will see in a moment.

\BeginLemma{repr-pdl}
$
\dasL{A}{\tau}{q} = 
\{s \in S \mid K_\tau(s)(A) \leq q\}
$
holds for all programs $\tau$, all measurable sets $A\in\Borel{S}$ and all $q\in[0, 1]$.
\EndLemma

\BeginProof
1. The proof proceeds by induction on $\tau$. Assume that $\tau = \gamma\in\Gamma$ is a primitive program, then
\begin{align*}
K_\gamma(s)(A) \leq q
& \Leftrightarrow
K_\gamma(s)\notin\bas{A}{> q} \\
& \Leftrightarrow
\bas{A}{> q} \notin P_\gamma(s) \\
& \Leftrightarrow
s \notin \dasI{A}{\tau}{q}
\end{align*}

2.
Assume that the assertion is true for $\tau_1$ and $\tau_2$, then
\begin{align*}
\dasL{A}{\tau_1\cup\tau_2}{q}
& = \bigcap_{\langle a_1, a_2\rangle\in Q^{(k)}(q)}\bigl(\dasL{A}{\tau_1}{a_1}\cap\dasL{A}{\tau_2}{a_2}\bigr)\\
& = \bigcap_{\langle a_1, a_2\rangle\in Q^{(k)}(q)}\bigl(\{s \mid K_{\tau_1}(s)(A) \leq a_1\}\cap\{s \mid K_{\tau_2}(s)(A) \leq a_2\}\bigr)\\
& = \{s \in S \mid (K_{\tau_1} + K_{\tau_2})(s)(A) \leq q\}\\
& = \{s \in S \mid K_{\tau_1\cup\tau_2}(s)(A) \leq q\}
\end{align*}

3.
The proof for angelic iteration $\tau^*$ is very similar, observing that 
$
\sum_{n \geq 0} K_{\tau^n}(s)(A) \leq q
$
iff there exists a sequence $\Folge{a}\in Q^{(\infty)}(q)$ with $K_{\tau^n}(s)(A) \leq a_n$ for all $n\in\Nat$.

4.
Finally, assume that the assertion is true for program $\tau$, and take $\gamma\in\Gamma$. Then, borrowing the notation from~(\ref{def_g})
\begin{align*}
G_\tau(A, q) \notin P_\gamma(s) 
& \Leftrightarrow
K_\gamma(s)\notin G_\tau(A, q)\\
& \Leftrightarrow
\int_0^1 K_\gamma(s)(\dasI{A}{\tau}{r})\ dr \leq q\\
& \stackrel{(\dagger)}{\Leftrightarrow}
\int_0^1 K_\gamma(s)(\{t \in S \mid K_\tau(t)(A) > r\})\ dr \leq q\\
& \stackrel{(\ddagger)}{\Leftrightarrow}
\int_S K_\tau(t)(A)\ K_\gamma(s)(dt) \leq q\\
& \stackrel{(\star)}{\Leftrightarrow}
K_{\gamma;\tau}(s)(A) \leq q.
\end{align*}
Here $(\dagger)$ is the induction hypothesis, $(\ddagger)$ derives from Choquet's Theorem~\ref{Choquet} and $(\star)$ comes from the definition of the convolution. 
\EndProof

It follows from this representation that for each program $\tau$ the set 
$
\dasL{A}{\tau^*}{q}
$
is a measurable subset of $S$, provided $A \in \Borel{S}$. This holds even without the assumption that the state space $S$ is closed under the Souslin operation. The latter assumption was made in~\cite{EED-PDL-TR}.

\BeginProposition{p-repr-pdl}
If games $\tau_1, \tau_2, \tau_3$ are in the PDL fragment, and the game frame $\Ge$ is generated by a Kripke frame, then 
\begin{align}
\label{l-distr}
\dasI{A}{\tau_1;(\tau_2\cup\tau_3)}{q} & = \dasI{A}{\tau_1;\tau_2\cup\tau_1;\tau_3}{q}\\
\label{r-distr}
\dasI{A}{(\tau_1\cup\tau_2);\tau_3}{q} & = \dasI{A}{\tau_1;\tau_3\cup\tau_2;\tau_3}{q}
\end{align}
for all $A \in \Borel{S}, q \geq 0$. 
\EndProposition

\BeginProof
Right distributivity~(\ref{r-distr}) is a basic assumption, which is given here for the sake of completeness. It remains to establish left distributivity~(\ref{l-distr}). Here it suffices to prove the equality for the respective complements. But this is easily established through Lemma~\ref{repr-pdl} and the observation that 
$
K_{\tau_1;(\tau_2\cup\tau_3)} = K_{\tau_1;\tau_2} + K_{\tau_1;\tau_3}
$
holds, because integration of non-negative functions is additive.
\EndProof

\section{Game Models}
\label{sec:game-models}

Game logic is a modal logic where the modalities are given through
games; it is defined through grammar
\begin{equation*}
  \phi = \top~\mid~p~\mid~\phi_1\wedge\phi_2~\mid~\langle \tau \rangle_q \phi,
\end{equation*}
see~(\ref{mod-grammar}). Here $p\in \AE$ is an atomic proposition, $\tau$ is a game, and $q\in
[0, 1]$ is a real number. Intuitively, formula $\langle \tau \rangle_q
\phi$ is true in state $s$ if playing game $\tau$ in state $s$ will
result in a state in which formula $\phi$ holds with a probability
greater than $q$.

\BeginDefinition{game-model}
A \emph{game model} $\Ge = (S, (P_\gamma)_{\gamma\in\Gamma}, (V_p)_{p\in \AE})$
over measurable space $S$ is given by a game frame $ (S, (P_\gamma)_{\gamma\in\Gamma})$, and by
a family $(V_p)_{p\in \AE}$ of sets which assigns to each atomic
statement a measurable set of state space $S$. We denote the
underlying game frame by $\Ge$ as well.
\EndDefinition
Define the validity sets for each formula recursively as follows:
\begin{align*}
\Gilt[\top]{\Ge} & := S\\
\Gilt[p]{\Ge} & := V_p, \text{ if } p \in \AE\\
\Gilt[\phi_1\wedge\phi_2]{\Ge} & :=   \Gilt[\phi_1]{\Ge}\cap\Gilt[\phi_2]{\Ge}\\
\Gilt[\langle \tau \rangle_q \phi]{\Ge} & := \dasI{\Gilt{\Ge}}{\tau}{q}
\end{align*}

Accordingly, we say that formula $\phi$ holds in state $s$ ($\Ge, s \models \phi$) iff $s \in \Gilt{\Ge}$.

The definition of $\Gilt[\langle \tau \rangle_q \phi]{\Ge}$ has a coalgebraic flavor. Coalgebraic logics define the validity of modal formulas through special natural transformations (called \emph{predicate liftings}) associated with the modalities~\cite{Moss,Schroeder-Expressivity,Schubert-ISDT}. This connection becomes manifest through Example~\ref{nat-transf} where $\dasI{\cdot}{\tau}{q}$ is shown to be a natural transformation. We will not, however, pursue this general approach further in this paper, since stochastic effectivity functions pose their own specific problems.
   
\BeginProposition{gilt-is-meas}
If state space $S$ is closed under the Souslin operation, $\Gilt{\Ge}$ is a
measurable subset for all formulas $\phi$. Moreover, 
$
\{\langle s, r\rangle \mid s \in \Gilt[\langle \tau \rangle_r \phi]{\Ge}\}
\in \Borel{S\otimes[0, 1]}.
$
\EndProposition

\BeginProof
The proof proceeds by induction on the formula $\phi$. If $\phi = p
\in \AE$ is an atomic proposition, then the assertion follows from $V_p
\in \Borel{S}$. The induction step uses Proposition~\ref{prop-interp}.
\EndProof

Stochastic Kripke models are defined similarly to game models: $\Ke = (S, (K_\gamma)_{\gamma\in\Gamma}, (V_p)_{p \in \AE})$ is called a \emph{stochastic Kripke model} iff $(S, (K_\gamma)_{\gamma\in\Gamma})$ is a stochastic Kripke frame with $V_p\in\Borel{S}$ for each atomic proposition $p \in \AE$. Validity of a formula in the state of a stochastic Kripke model is defined as validity in the associated game model. Thus we know that for primitive games $\gamma_1, \dots, \gamma_n\in \Gamma$
\begin{equation}
\label{kri-rep}
s \in \Gilt[\langle \gamma_1;\dots;\gamma_n\rangle_q \phi]{\Ge}
\Leftrightarrow
\Ke, s \models \langle \gamma_1;\dots;\gamma_n\rangle_q \phi
\Leftrightarrow
\bigl(K_{\gamma_1}\star\dots\star K_{\gamma_n}\bigr)(s)(\Gilt{\Ke}) \geq q
\end{equation}
holds (Proposition~\ref{primitive-product}), and that games are semantically equivalent to their distributive counterparts (Proposition~\ref{p-repr-pdl}).

One of the corner stones for the interpretation of model logics through a stochastic Kripke model $\Ke$ is the observation that the validity sets $\Gilt{\Ke}$ for formulas $\phi$ is measurable~\cite{Panangaden-book,EED-CoalgLogic-Book}. Proposition~\ref{gilt-is-meas} together with~(\ref{kri-rep}) gives a much more general result by stating that 
$
\{\langle s, r\rangle \mid s \in \Gilt[\langle \tau \rangle_r \phi]{\Ke}\}
$
is actually measurable in the product space $S\otimes[0, 1]$ (from which the original statement may be obtained by taking cuts). This is valid in general measurable spaces without the Souslin condition. 

\medskip

We assume from now on that the state spaces of our models are closed under the Souslin operation. 

\subsection{Morphisms and Congruences}
\label{sec:mod-morph}

Let $\He = (T, (Q_\gamma)_{\gamma\in\Gamma}, (W_p)_{p\in \AE})$ be a
second game model, then a measurable map $f: S \to T$ which is also a
frame morphism $f: (S, (P_\gamma)_{\gamma\in\Gamma}) \to (T,
(Q_\gamma)_{\gamma\in\Gamma})$ is called a \emph{model morphism} $f:
\Ge\to\He$ iff $ \InvBild{f}{W_p} = V_p $ holds for all atomic
propositions, i.e., if $f(s)\in W_p$ iff $s\in V_p$ always
holds. Model morphisms are compatible with validity:

\BeginProposition{mod-pres}
Let $\phi$ be a formula of game logic, $f: \Ge\to\He$ be a model morphism. Then 
\begin{equation*}
  \Ge, s \models \phi \text{ iff }  \He, f(s) \models \phi. 
\end{equation*}
\EndProposition

\BeginProof
The claim is equivalent to saying that
\begin{equation*}
  \Gilt{\Ge} = \InvBild{f}{\Gilt{\He}}
\end{equation*}
for all formulas $\phi$. This is established through induction on the
formula $\phi$. Because $f$ is a model morphism, the assertion holds
for atomic proposition. The induction step is established through
Proposition~\ref{g-morph-compatible}.
\EndProof

An equivalence relation $\rho$ on the state space of a game model $\Ge = (S, (P_\gamma)_{\gamma\in\Gamma}, (V_p)_{p\in \AE})$ is said to be a \emph{congruence for $\Ge$} iff
\begin{enumerate}
\item $\isEquiv{s}{s'}{\rho}$ implies $s \in V_p \Leftrightarrow s'\in V_p$ for all atomic sentences $p\in \AE$ (equivalently, iff $V_p\in\Inv{\rho}{S}$ for all $p \in \AE$),
\item $\rho$ is a congruence for  effectivity function $P_\gamma$ for all $\gamma\in\Gamma$. 
\end{enumerate}
 
As an illustration, assume that $f: \Ge\to\He$ is a model morphism
such that $f\times id_{[0, 1]}$ constitutes also a final map, then $\Kern{f}$ is a congruence
for $\Ge$. In fact, $\Kern{f}$ constitutes a congruence for each
effectivity function by Proposition~\ref{final-is-congr}. Because
$\InvBild{f}{W_p} = V_p$ holds for each atomic proposition $p$ (with
$W_p$ as the sets in which $p$ holds in model $\He$), we conclude
that each $V_p$ in $\Kern{f}$-invariant.

Given a congruence for $\Ge$, we construct the factor model $\Faktor{\Ge}{\rho}$ in a straightforward manner: Let $Q_\gamma$ be the effectivity function on $\Faktor{S}{\gamma}$ associated with $P_\gamma$, then 
\begin{equation*}
\Faktor{\Ge}{\rho} := (\Faktor{S}{\rho}, (Q_\gamma)_{\gamma\in\Gamma}, (\Bild{\fMap{\rho}}{V_p})_{p\in \AE})
\end{equation*}
is a game model with $\fMap{\rho}:\Ge\to\Faktor{\Ge}{\rho}$ as a morphism. 

A congruence for a stochastic Kripke model is defined in exactly the same way: The extensions for the atomic propositions are assumed to be invariant, and the equivalence is a congruence for each stochastic relation. Corollary~\ref{stochCongr-is-effCongr} then shows that a congruence for a Kripke models is also a congruence for the associated game model. 

\subsection{The Equivalence Induced by the Logic}
\label{sec:induced}

We investigate now the equivalence relation $\rho$ induced by the logic on the state space. Thus two states are equivalent iff they satisfy exactly the same formulas. Formally, define the \emph{theory of a state} as all formulas which the state satisfies, i.e., 
$
\theTheory{\Ge}{s} := \{\phi \mid \Ge, s \models \phi\}.
$
This induces an equivalence relation on state space $S$ through
\begin{equation*}
  \isEquiv{s}{s'}{\rho} \text{ iff } 
\theTheory{\Ge}{s} = \theTheory{\Ge}{s'},
\end{equation*}
so $\isEquiv{s}{s}{\rho}$ iff states $s$ and $s'$ satisfy exactly the same formulas. It is clear that the validity sets $\Gilt{\Ge}$ are $\rho$-invariant. We make the assumption that the validity sets generate the $\sigma$-algebra of invariant measurable sets.
\BeginDefinition{s-small}
The state space $S$ of model $\Ge$ is said to be \emph{small} iff $\rho$ is tame so that 
$\rho$ is exact with $\sigma(\{\Gilt{\Ge} \mid \phi\text{ is a formula}\})$. Model $\Ge$ itself is said to be \emph{small} iff its state space is small.
\EndDefinition

Because each validity set $\Gilt{\Ge}$ is $\rho$-invariant and
measurable by Proposition~\ref{gilt-is-meas}, we know  that the
$\sigma$-algebra $ \sigma(\{\Gilt{\Ge} \mid \phi\text{ is a
  formula}\}) $ is always contained in $\Inv{\rho}{S}$. Smallness then implies
that the former $\sigma$-algebra exhausts the invariant sets. This latter
assumption is satisfied, e.g., in case the state space is Polish or
analytic, and there are only countably many
formulas. By~\cite[Corollary 2.6.5]{EED-CoalgLogic-Book}, the condition
\begin{equation}
\label{its-small}
  \Inv{\rho}{S} = \sigma(\{\Gilt{\Ge} \mid \phi\text{ is a formula}\}).
\end{equation}
is equivalent so saying that
\begin{equation*}
  \Borel{\Faktor{S}{\rho}} = \sigma(\{A \subseteq \Faktor{S}{\rho} \mid \InvBild{\fMap{\rho}}{A} = \Gilt{\Ge}\text{ for some formula } \phi\}).
\end{equation*}
\cite[Example 2.6.7]{EED-CoalgLogic-Book} shows that there are measurable spaces
which do not satisfy (\ref{its-small}) (and for which,
accordingly, the equivalence induced by the logic ---~in that case a simple
negation free Hennessy Milner logic~--- is not a congruence).

The model is assumed to satisfy a special condition which is intended
as a technical condition preventing that different states show widely
diverging behavior.
\BeginDefinition{Frege-cond}
Game frame $\Ge$ is said to satisfy the \emph{Frege condition}%
given states $s$ and $s'$ and a simple game $\gamma\in \Gamma$, then
set
\begin{equation*}
   \{A \in \Borel{\SubProb{S}} \mid A \in P_\gamma(s) \Leftrightarrow A \in P_{\gamma}(s')\}
\end{equation*}
is closed under taking $\sigma$-algebras. 
\EndDefinition

Thus if we know that 
$
A\in P_\gamma(s) \Leftrightarrow A\in P_\gamma(s')
$
for $A \in \mathcal{A}$, then we may conclude that this property holds also for all $A \in \sigma(\mathcal{A})$. We say that a \emph{game model has the Frege property} iff its underlying frame enjoys it.

Kripke generated frames satisfy actually a stronger condition: It is not difficult to
see that for each stochastic relation $K: S \Trans S$ the set
\begin{equation}
\label{a-frege}
   \{A \in \Borel{\SubProb{S}} \mid K(s)\in A \Leftrightarrow K(s')\in A\} 
\end{equation}
is in fact a $\sigma$-algebra itself. 

We assume for the rest of this section that the state space $S$ of the model is small, and that the model itself satisfies the Frege condition.

\BeginLemma{small-for-subprob}
$\bS{\Faktor{S}{\rho}}$ is generated by
$
\{\bas{\Bild{\fMap{\rho}}{\Gilt{\Ge}}}{> q} \mid q \geq 0, \phi\text{ is a formula}\},
$
provided $S$ is small.
\EndLemma

\BeginProof
It is easy to establish that 
$
\Borel{\SubProb{T}} = \sigma(\{\bas{A}{> q} \mid q \geq 0, A \in {\cal A}\})
$ 
if we know that 
$
\Borel{T} = \sigma({\cal A}).
$
\EndProof

Smallness gives an interesting relationship between spaces of subprobabilities: The condition on equality of Borel sets translates into a Borel isomorphism between the corresponding spaces of subprobabilities.

\BeginLemma{borel-isom}
$\SubProb{\Inv{\rho}{S}}$ is Borel isomorphic to $\SubProb{\Faktor{S}{\rho}}$.
\EndLemma

\BeginProof
1.
Define
\begin{equation*}
\hbar:
\begin{cases}
 \SubProb{\Inv{\rho}{S}} &\to  \SubProb{\Faktor{S}{\rho}}, \\
  \mu     &\mapsto  \SubProb{\fMap{\rho}}(\mu).
\end{cases}
\end{equation*}
Assume that $\hbar(\mu)(D) = \hbar(\mu')(D)$ for all $D \in \Borel{\Faktor{S}{\rho}}$, then we conclude from the $\pi$-$\lambda$-Theorem~\ref{pi-lambda} that $\mu(B) = \mu'(B)$ for all $B \in \Inv{\rho}{S}$. In fact, define 
\begin{equation*}
\mathcal{D} := \{A \in \Inv{\rho}{S}\mid \mu(A) = \mu'(A)\},
\end{equation*}
then $\mathcal{D}$ is closed under complementation and countable disjoint unions; moreover, $S\in\mathcal{D}$. If 
$
E := \Gilt{\Ge}
$
for some formula $\phi$, then $\Bild{\fMap{\rho}}{E}\in\Borel{\Faktor{S}{\rho}}$, and
\begin{equation*}
\mu(E) = \hbar(\mu)(\Bild{\fMap{\rho}}{E}) = \hbar(\mu')(\Bild{\fMap{\rho}}{E}) = \mu'(E).
\end{equation*}
Thus 
$
\mathcal{E} := \{\Gilt{\Ge} \mid \phi\text{ is a formula}\} \subseteq \mathcal{D}.
$
Since $S$ is small, $\sigma(\mathcal{E}) = \mathcal{D}$, and since the logic is closed under conjunctions, $\mathcal{E}$ is closed under finite intersections. Thus $\mu = \mu'$ by Theorem~\ref{pi-lambda}. Consequently, $\hbar$ is injective. 

Let $\nu\in\SubProb{\Faktor{S}{\rho}}$, and define $\nu_0(A) := \nu(\Bild{\fMap{\rho}}{A})$ for $A \in \Inv{\rho}{A}$. Because $A = \InvBild{\fMap{\rho}}{\Bild{\fMap{\rho}}{A}}$ for $\rho$-invariant $A\in\Borel{S}$, we conclude that $\Bild{\fMap{\rho}}{A}\in\Borel{\Faktor{S}{\rho}}$, so that $\nu_0$ is defined on all of $\Inv{\rho}{S}$. $\nu_0$ is monotone, and is additive because $\Bild{\fMap{\rho}}{A\cup B} = \Bild{\fMap{\rho}}{A}\cup\Bild{\fMap{\rho}}{B}, \Bild{\fMap{\rho}}{A\cap B} = \Bild{\fMap{\rho}}{A}\cap\Bild{\fMap{\rho}}{B}$ for $\rho$-invariant sets $A$ and $B$. If $A_n \in \Inv{\rho}{S}$ is a decreasing sequence with $A_1 \supseteq \dots \supseteq A_n \supseteq\dots $ and 
$
\bigcap_{n\in\Nat} A_n = \emptyset,
$
we know that the sequence $B_n$ with $B_n := \Bild{\fMap{\rho}}{A_n}$ decreases with 
$
\bigcap_{n\in\Nat} B_n = \emptyset,
$
because $\fMap{\rho}^{-1}$ is injective, and
$
A_n = \InvBild{\fMap{\rho}}{B_n}.
$
Thus $\nu_0$ is $\sigma$-additive, hence $\nu_0\in\SubProb{\Inv{\rho}{S}}$ with plainly $\hbar(\nu_0) = \nu$. Thus $\hbar$ is a bijection.

2.
Because
\begin{align*}
\InvBild{\hbar}{\basS{\Faktor{S}{\rho}}{B}{> q}} & = 
\basS{\Inv{\rho}{S}}{\InvBild{\fMap{\rho}}{B}}{> q},\\
\Bild{\hbar}{\basS{\Inv{\rho}{S}}{A}{> q}} & = 
\basS{\Faktor{S}{\rho}}{\Bild{\fMap{\rho}}{A}}{> q},
\end{align*}
we conclude that both the image and the inverse image of measurable sets is measurable.
\EndProof

Put, as above,  
\begin{equation}
\label{g-tau}
G_\tau(\Gilt{\Ge}, q) := \{\mu \in \SubProb{S} \mid \int_0^1 \mu(\Gilt[\langle\tau\rangle_r\phi]{\Ge})\ dr > q\}
\end{equation}
for game $\tau$ and formula $\phi$ and $q \geq 0$. Then evidently
\begin{equation}
\label{beta-G}
\bas{\Gilt{\Ge}}{> q} = G_\epsilon(\Gilt{\Ge}, q).
\end{equation}
Moreover, putting
\begin{equation*}
  H := \{\nu \in \SubProb{\Faktor{S}{\rho}} \mid \int_0^1 \nu(\Bild{\fMap{\rho}}{\Gilt[\langle\tau\rangle_r\phi]{\Ge}})\ dr > q\},
\end{equation*}
we see that
\begin{equation*}
  G_\tau(\Gilt{\Ge}, q) = \InvBild{\SubProb{\fMap{\rho}}}{H},  
\end{equation*}
and that 
$
H \in \bS{\Faktor{S}{\rho}}.
$
Thus
\begin{align*}
  \sigma(\{G_\tau(\Gilt{\Ge}, q) \mid \phi\text{ is a formula}, q \geq 0\})
& \subseteq 
 \InvBild{\SubProb{\fMap{\rho}}}{\bS{\Faktor{S}{\rho}}} \\
& \stackrel{(\dagger)}{=}
\sigma(\{\bas{\Gilt{\Ge}}{q} \mid  \phi\text{ is a formula}, q \geq 0\})\\
& = 
\sigma(\{G_\epsilon(\phi, q) \mid  \phi\text{ is a formula}, q \geq 0\}).
\end{align*}
Equation~$(\dagger)$ follows from Lemma~\ref{small-for-subprob} with (\ref{beta-G}). Consequently, the sets
$G_\tau(\Gilt{\Ge}, q)$ generate the $\sigma$-algebra $\InvBild{\SubProb{\fMap{\rho}}}{\bS{\Faktor{S}{\rho}}}$.
 
Now assume that $s$ and $s'$ satisfy exactly the same formulas, thus $\isEquiv{s}{s'}{\rho}$, then we have for the elementary game $\gamma\in \Gamma$
\begin{align*}
G_\tau(\Gilt{\Ge}, q) \in P_\gamma(s)
&\Leftrightarrow 
\Ge, s \models \langle\gamma;\tau\rangle_q\phi \\
&\Leftrightarrow
\Ge, s' \models \langle\gamma;\tau\rangle_q\phi \\
&\Leftrightarrow
G_\tau(\Gilt{\Ge}, q) \in P_\gamma(s').
\end{align*}
Because $\Ge$ satisfies the Frege condition, we infer 
\begin{equation*}
\InvBild{\SubProb{\fMap{\rho}}}{\bS{\Faktor{S}{\rho}}} \subseteq \{ A \in \Borel{\SubProb{S}} \mid A \in P_\gamma(s) \Leftrightarrow A \in P_\gamma(s')\}.
\end{equation*}

Thus we have shown
\BeginProposition{rho-is-congr}
Assume that the game model satisfies the Frege condition, and that it has a small state space, then the equivalence relation induced by game logic is a congruence on $\Ge$, its factor map is a morphism. 
\EndProposition

\BeginProof
The above argumentation shows through Proposition~\ref{is-a-congruence} that $\rho$ is a congruence for each effectivity function $P_\gamma$. This is so since we observe for $\isEquiv{s}{s'}{\rho}$ and for $H \in \bS{\Faktor{S}{\rho}}$ the equivalence 
\begin{equation*}
\InvBild{\fMap{\rho}}{H} \in P_\gamma(s) \Leftrightarrow \InvBild{\fMap{\rho}}{H} \in P_\gamma(s').
\end{equation*}
It is plain that $V_p\in\Inv{\rho}{S}$ for each atomic statement $p$. 
\EndProof

Because stochastic Kripke models satisfy the Frege condition, we obtain as a consequence
\BeginCorollary{K-rho-is-congr}
Let $\Ke$ be a stochastic Kripke model over a small state space. Then the equivalence relation induced by the game logic is a congruence for the game model associated with $\Ke$.
\QED
\EndCorollary

We have a look now at logical equivalence of two models. Fix in addition to model $\Ge$ another model $\He = (T, (Q_\gamma)_{\gamma\in\Gamma}, (W_p)_{p\in \AE})$, and assume that $\He$ is small and satisfies the Frege condition. Let $\rho$ be the equivalence on the state space of $\Ge$ induced by the logic, and let $\theta$ be its counterpart on the state space of $\He$. Then both equivalence relations are congruences by Proposition~\ref{rho-is-congr}, hence both factor models $\Faktor{\Ge}{\rho}$ and $\Faktor{\He}{\theta}$ are defined. 

Logical equivalence of models is defined in terms of the theories of states:
\BeginDefinition{log-equiv}
Models $\Ge$ and $\He$ are called \emph{logically equivalent} iff 
\begin{equation*}
\{\theTheory{\Ge}{s} \mid s \in S\} = \{\theTheory{\He}{t} \mid t \in T\}.
\end{equation*}
\EndDefinition
Thus in logically equivalent models each state in one models finds a state in the other model with exactly the same theory.

Now assume that the factor spaces $\Faktor{\Ge}{\rho}$ and $\Faktor{\He}{\theta}$ are isomorphic, then it is plain that the models are logically equivalent. The interesting part is the converse. 

Let $\Ge$ and $\He$ be logically equivalent, and define 
\begin{equation*}
R := \{\langle\Klasse{s}{\rho}, \Klasse{t}{\theta}\rangle \mid s \in S, t \in T\}.
\end{equation*}

\BeginLemma{ist-graph}
$R$ is the graph of a bijection $\alpha: \Faktor{S}{\rho}\to\Faktor{T}{\theta}$; $\alpha$ is bi-measurable. 
\EndLemma

\BeginProof
\cite{EED-CoalgLogic-Book}, Lemma 2.6.10 together with Corollary 2.6.5.
\EndProof

Before investigating the Borel isomorphism $\alpha$ further, the Frege condition is extended to classes of frames.

\BeginDefinition{Frege-classes}
A class of game frames is said to be a \emph{Frege class} iff the following holds: whenever $(S, (P_\gamma)_{\gamma\in\Gamma})$ and $(T, (Q_\gamma)_{\gamma\in\Gamma})$ are frames in this class, and whenever 
\begin{equation*}
\SubProb{S}
\stackrel{g}{\longrightarrow}
\SubProb{M}
\stackrel{h}{\longleftarrow}
\SubProb{T}
\end{equation*}
are maps for some witness space $M$, then the set 
\begin{equation*}
\{V\in\Borel{\SubProb{M}} \mid \InvBild{g}{V} \in P_\gamma(t) \Leftrightarrow \InvBild{h}{V}\in Q_\gamma(s)\}
\end{equation*}
is closed under forming $\sigma$-algebras ($s\in S, t \in T, \gamma\in\Gamma$). 
\EndDefinition

Thus maps $g$ and $h$ between the state spaces ties the frames loosely together (note that we do not assume that neither is measurable). This property is extended to models in the obvious way.
 
Again it is plain that the class of Kripke generated frames with respective stochastic relations $K_\gamma: S \Trans S$ and $L_\gamma: T \Trans T$ is a Frege class. This is so since the set
\begin{equation*}
\{A \in \Borel{\SubProb{T}} \mid L_\gamma(t) \in \InvBild{h}{A} \Leftrightarrow K_\gamma(s)\in \InvBild{g}{A}\}
\end{equation*}
is a $\sigma$-algebra. 

\BeginProposition{alpha-is iso}
Assume that $\Ge$ and $\He$ are taken from a Frege class of models. Then 
$
\alpha: \Faktor{\Ge}{\rho}\to\Faktor{\He}{\theta}
$
is an isomorphism.
\EndProposition
{
\def\Pt{\widetilde{P}}
\def\Qt{\widetilde{Q}}
\def\Vt{\widetilde{V}}
\def\Wt{\widetilde{W}}
\BeginProof
0.
We show that $\alpha: \Faktor{\Ge}{\rho}\to\Faktor{\He}{\theta}$ is a morphism, interchanging the r{\^o}les of $\Ge$ and $\He$ and using the same arguments shows that $\alpha^{-1}$ is a morphism as well.

1.
Let $\Pt_\gamma: \Faktor{\Ge}{\rho}\to \Vau{\Faktor{\Ge}{\rho}}$ and
$\Qt_\gamma: \Faktor{\He}{\theta}\to \Vau{\Faktor{\He}{\theta}}$ be the effectivity functions associated with the factor models, then we have to show that this diagram commutes for each game $\gamma\in\Gamma$
\begin{equation*}
\xymatrix{
\Faktor{S}{\rho}\ar[rr]^{\alpha}\ar[d]_{\Pt_{\gamma}}&&\Faktor{T}{\theta}\ar[d]^{\Qt_{\gamma}}\\
\Vau{\Faktor{S}{\rho}}\ar[rr]_{\Vau{\alpha}}&&\Vau{\Faktor{T}{\theta}}
}
\end{equation*}
Consequently, we have to show that 
\begin{equation}
\label{eqq}
\InvBild{\SubProb{\fMap{\theta}}}{W}\in Q_\gamma(t) \Leftrightarrow \InvBild{\SubProb{\alpha\circ\fMap{\rho}}}{W}\in P_\gamma(s)
\end{equation}
whenever 
$
\langle\Klasse{s}{\rho}, \Klasse{t}{\theta}\rangle\in R 
$
and 
$
W \in \bS{\Faktor{T}{\theta}}.
$

2.
In fact, let 
$
W = \basS{\Faktor{T}{\theta}}{\Bild{\fMap{\theta}}{\Gilt{\He}}}{q},
$
then 
\begin{align*}
\InvBild{\SubProb{\fMap{\theta}}}{W} 
& = 
\basS{T}{\Gilt{\He}}{q},\\
\InvBild{\SubProb{\alpha\circ\fMap{\rho}}}{W} 
& = 
\basS{S}{\Gilt{\Ge}}{q}.
\end{align*}
Since 
$
\theTheory{\Ge}{s} = \theTheory{\He}{t},
$
we conclude
\begin{align*}
\basS{T}{\Gilt{\He}}{q} \in Q_\gamma(t) 
&\Leftrightarrow
\He, t \models \langle\gamma\rangle_q\phi\\
&\Leftrightarrow
\Ge, s \models \langle\gamma\rangle_q\phi\\
&\Leftrightarrow
\basS{S}{\Gilt{\Ge}}{q} \in P_\gamma(s) 
\end{align*}
Thus we have established the equivalence~(\ref{eqq}) for 
$
W = \basS{\Faktor{T}{\theta}}{\Bild{\fMap{\theta}}{\Gilt{\He}}}{q},
$
so from the observation in Lemma~\ref{small-for-subprob} we see that 
\begin{equation*}
\sigma(\{\basS{\Faktor{T}{\theta}}{\Bild{\fMap{\theta}}{\Gilt{\He}}}{q}\}\mid
\phi\text{ is a formula}, q \geq 0\}) = \bS{\Faktor{T}{\theta}}.
\end{equation*}
From the Frege condition we conclude now that the diagram above commutes. 

3.
We have finally to show that 
$
\InvBild{\alpha}{\Wt_p} = \Vt_p,
$
where $\Vt_p$ resp. $\Wt_p$ is the respective of atomic proposition $p\in \AE$. This is obvious.
\EndProof
}

\BeginDefinition{beh-equiv}
Models $\Ge$ and $\He$ are called \emph{behaviorally equivalent} iff 
\begin{equation*}
\xymatrix{
\Ge\ar[r]^f&\Me&\He\ar[l]_g
}
\end{equation*}
for some model $\Me$ and surjective morphisms $f, g$.
\EndDefinition
Thus we can find in the situation above for each state $s$ of $\Ge$ a state $t$ of $\He$ such that $f(s) = g(t)$, and vice versa, which entails that in $s$ and in $t$ exactly the same formulas hold. 
We summarize the discussion:
\BeginProposition{express}
Let $\Ge$ and $\He$ be models, and consider these statements
\begin{enumerate}[a.]
\item\label{express-1} $\Ge$ and $\He$ are behaviorally equivalent.
\item\label{express-2} $\Ge$ and $\He$ are logically equivalent. 
\end{enumerate}
Then we have
\begin{enumerate}[i.]
\item\label{express-i} $\labelImpl{express-1}{express-2}$ always holds.
\item\label{express-ii} $\labelImpl{express-2}{express-1}$ holds, provided the models are small and are taken from a Frege class of models.
\end{enumerate}
\EndProposition

\BeginProof
1.
Let $\Ge$ and $\He$ be behaviorally equivalent, and take model $\Me$ and morphisms $f, g$ as in Definition~\ref{beh-equiv}. Let $s\in S$, then there exists by surjectivity $t\in T$ with $f(s) = g(t)$, so plainly by Proposition~\ref{mod-pres}
\begin{equation*}
\theTheory{\Ge}{s} = \theTheory{\Me}{f(s)} = \theTheory{\Me}{g(t)} = \theTheory{\He}{t}. 
\end{equation*}
Consequently, $\Ge$ and $\He$ are logically equivalent.

2.
Let $\rho$ and $\theta$ be the equivalence relations associated with the logic, then we have
\begin{equation*}
\xymatrix{
\Ge\ar[dr]_{\fMap{\rho}} && \He\ar[dl]^{\fMap{\theta}}\\
&\Faktor{\Ge}{\rho} \cong \Faktor{\He}{\theta}&}
\end{equation*}
with $\cong$ as the isomorphism according to Proposition~\ref{alpha-is iso}. We know from Proposition~\ref{rho-is-congr} that the factor maps are surjective morphisms. 
\EndProof

If the models are generated from stochastic Kripke models, we can say a little bit more.
\BeginCorollary{express-stoch}
Let $\Ge$ and $\He$ be models which are obtained from stochastic Kripke models, and consider these statements
\begin{enumerate}[a.]
\item\label{express-a1} $\Ge$ and $\He$ are behaviorally equivalent.
\item\label{express-a2} $\Ge$ and $\He$ are logically equivalent. 
\end{enumerate}
Then we have
\begin{enumerate}[i.]
\item\label{express-ai} $\labelImpl{express-a1}{express-a2}$ always holds.
\item\label{express-aii} If the models are small, then the statements above are equivalent. 
\end{enumerate}
\EndCorollary

\BeginProof
This follows from Proposition~\ref{express} together with the observation that the class of stochastically generated models is a Frege class. 
\EndProof

\subsection{The Test Operator}
\label{sec:test-op}

The test operator has not been incorporated into the discussions so far. Given a formula $\phi$, Angel may test whether or not the formula is satisfied; this yields the two games $\posTest$ and $\negTest$. Game $\posTest$ checks whether formula $\phi$ is satisfied in the current state; if it is, Angel continues with the next game, if it is not, Angel loses. Similarly for $\negTest$: Angel checks, whether formula $\phi$ is not satisfied. Note that we do not have negation in our logic, so we cannot test for $\neg\phi$. We can test, however, whether a state state $s$ does not satisfy a formula $\phi$ by evaluating $s \in S \setminus \Gilt{\Ge}$, since complementation is available in our $\sigma$-algebra. 

In order to seamlessly integrate these testing games into our models, we define for each formula two effectivity functions for positive and for negative testing, resp. The following technical observations will be helpful.

\BeginLemma{make-effFnct}
Let $P$ be a stochastic effectivity function on $S$, and assume that $F: \SubProb{S}\to\SubProb{S}$ is measurable, then
$ 
P'(s) := \{W \in \Borel{\SubProb{S}} \mid \InvBild{F}{W} \in P(s)\}
$
defines a stochastic effectivity function on $S$.
\EndLemma

\BeginProof
$P'(s)$ is upward closed, since $P(s)$ is, so t-measurability has to be established. Let 
$
H \in \Borel{\SubProb{S}\otimes[0, 1]}
$
be a test set, then 
$
H_q \in P'(s) \Leftrightarrow \bigl(\InvBild{(F\times id_{[0, 1]})}{H}\bigr)_q \in P(s).
$
Since $F$ is measurable, $F\times id_{[0, 1]}: \SubProb{S}\times[0, 1]\to\SubProb{S}\times[0, 1]$ is, hence 
$
\InvBild{(F\times id_{[0, 1]})}{H}
$
is a member of
$
\Borel{\SubProb{S}\otimes[0, 1]}.
$
Because $P$ is t-measurable, we conclude
$
\{\langle s, q\rangle \mid H_q \in P'(s)\}\in \Borel{S\otimes[0, 1]}.
$ 
\EndProof

\BeginLemma{test-op1}
Define for  $A \in \Borel{S}$, $\mu\in\SubProb{S}$ and $B \in \Borel{S}$ the localization to $A$ as
$
F_A(\mu)(B) := \mu(A\cap B).
$
Then $F_A: \SubProb{S}\to\SubProb{S}$ is measurable.
\EndLemma

\BeginProof
This follows from
$
\InvBild{F_A}{\bas{C}{\bowtie q}} = \bas{A \cap C}{\bowtie q}.
$
\EndProof

$F_A$ localizes measures to the set $A$ --- everything outside $A$ is discarded. 
Now define for state $s$ and formula $\phi$ 
\begin{align*}
P_{\posTest}(s) & := \{W \in \Borel{\SubProb{S}} \mid \InvBild{F_{\Gilt{\Ge}}}{W}\in I_D(s)\},\\
P_{\negTest}(s) & := \{W \in \Borel{\SubProb{S}} \mid \InvBild{F_{S \setminus \Gilt{\Ge}}}{W}\in I_D(s)\},
\end{align*}
where $I_D$ is the Dirac function defined in Example~\ref{dirac}. We obtain 

\BeginProposition{test-is-effFnct}
$P_{\posTest}$ and $P_{\negTest}$ define for each formula $\phi$ a stochastic effectivity function.
\EndProposition

\BeginProof
From Proposition~\ref{gilt-is-meas} we infer that $\Gilt{\Ge}\in \Borel{S}$, consequently, $F_{\Gilt{\Ge}}$ and $F_{S \setminus \Gilt{\Ge}}$ are measurable functions $\SubProb{S}\to\SubProb{S}$ by Lemma~\ref{test-op1}. Thus the assertion follows from Lemma~\ref{make-effFnct}. 
\EndProof

In~\cite[Section 6.5]{EED-PDL-TR} the integration of the test operators associated with PDL formulas is discussed.  Adapted to the current notation, the definitions for the associated stochastic relations $K_{\posTest}: S \Trans S$ and $K_{\negTest}: S \Trans S$ read
\begin{align*}
K_{\posTest}(s) & := F_{\Gilt{\Ge}}(D(s)),\\
K_{\negTest}(s) & := F_{S\setminus \Gilt{\Ge}}(D(s)).
\end{align*}
These relations can be defined for formulas of game logic as well. Thus we have, e.g., 
\begin{equation*}
K_{\posTest}(s)(B) = 
\begin{cases}
1, & \text{if } s \in B \text{ and } \Ge, s \models \phi \\
0, & \text{otherwise}.
\end{cases}
\end{equation*}
This is but a special case, since $P_{\posTest}$ and $P_{\negTest}$ are generated by these stochastic relations.

\BeginLemma{is-gen-test}
Let $\phi$ be a formula of game logic, then 
$
P_{\posTest} = P_{K_{\posTest}}
$
and
$
P_{\negTest} = P_{K_{\negTest}}.
$
\EndLemma

\BeginProof
The assertions follow from expanding the definitions.
\EndProof

This extension integrates well into the scenario, because it is compatible with morphisms. We will establish this now.

\BeginProposition{comp-morph}
Let $\Ge$ and $\He$ be game models over state spaces $S$ and $T$, resp. Assume that $f: \Ge\to\He$ is a morphism of game models; define for each formula $\phi$ the effectivity functions $P_{\posTest}$ and $P_{\negTest}$ for $\Ge$ resp. $Q_{\posTest}$ and $Q_{\negTest}$ for $\He$. Then $f$ is a morphism $P_{\posTest}\to Q_{\posTest}$ and $P_{\negTest}\to Q_{\negTest}$ for each formula $\phi$. 
\EndProposition

\BeginProof
Fix formula $\phi$; we will prove the assertion only for $\posTest$, the proof for $\negTest$ is the same. The notation is fairly overloaded. We will use primed quantities when referring to $\He$ and state space $T$, and unprimed ones when referring to model $\Ge$ with state space $S$.

Note first that 
$
\SubProb{f}(F_{\posTest}(D(s))) = F'_{\posTest}(D'(f(s))),
$
because we have for each $G \in \Borel{T}$
\begin{align*}
\SubProb{f}(F_{\posTest}(D(s)))(G)
& = 
F_{\posTest}(D(s))(\InvBild{f}{G})\\
& = 
D(s)(\Gilt{\Ge}\cap\InvBild{f}{G})\\
& \stackrel{(\dagger)}{=}
D(s)(\InvBild{f}{\Gilt{\He}}\cap\InvBild{f}{G})\\
& =
D'(f(s))(\Gilt{\He}\cap G) \\
& = 
F'_{\posTest}(D'(f(s)))(G).
\end{align*}
We have used
$
\InvBild{f}{\Gilt{\He}} = \Gilt{\Ge}
$
in ($\dagger$), since $f$ is a morphism, see Proposition~\ref{mod-pres}. But now we may conclude
\begin{equation*}
W \in \Vau{f}({P_{\posTest}(s)})
\Leftrightarrow
\SubProb{f}(F_{\posTest}(D(s))) \in W
\Leftrightarrow
F'_{\posTest}(D'(f(s))) \in W
\Leftrightarrow 
W \in Q_{\posTest}(f(s)),
\end{equation*}
hence 
$
\Vau{f}\circ P_{\posTest} = Q_{\posTest} \circ f
$
is established.
\EndProof

We compute $\Gilt[\langle{\posTest[p]};\tau\rangle_q\phi]{\Ge}$ and $\Gilt[\langle {\negTest[p]};\tau\rangle_q\phi]{\Ge}$ for a primitive formula $p\in\AE$ and an arbitrary game $\tau$ for the sake of illustration. 

First a technical remark: Let $\lambda$ be Lebesgue measure on the unit interval, then 
\begin{equation}
\label{lgth}
\Ge, s \models \langle\tau\rangle_q\phi \Leftrightarrow \lambda(\{r \in [0, 1] \mid \Ge, s \models \langle\tau\rangle_r\phi\}) > q
\end{equation}
In fact, the map $r \mapsto \Gilt[\langle\tau\rangle_r\phi]{\Ge}$ is monotone decreasing; this is intuitively clear: if Angel will have a strategy for reaching a state in which formula $\phi$ holds with probability at least $q > q'$, it will have a strategy for reaching such a state with probability at least $q'$. But this entails that the set $\{r \in [0, 1] \mid \Ge, s \models \langle\tau\rangle_r\phi\}$ constitutes an interval which contains 0 if it is not empty. This interval is longer than $q$ (i.e., its Lebesgue measure is greater than $q$) iff $q$ is contained in it. From this~(\ref{lgth}) follows.

Now assume $\Ge, s \models \langle{\posTest[p]};\tau\rangle_q\phi$. Thus 
\begin{equation*}
G_\tau(\Gilt{\Ge}, q) \in P_{\posTest[p]}(s) \Leftrightarrow 
F_{V_p}(D(s)) \in G_\tau(\Gilt{\Ge}, q) \Leftrightarrow 
\int_0^1D(s)(V_p\cap\Gilt[\langle\tau\rangle_r\phi]{\Ge})\ dr > q,
\end{equation*}
which means
\begin{equation*}
D(s)(V_p)\cdot\int_0^1D(s)(\Gilt[\langle\tau\rangle_r\phi]{\Ge})\ dr > q.
\end{equation*}

This implies 
\begin{equation*}
D(s)(V_p) = 1
\text{ and }
\int_0^1D(s)(\Gilt[\langle\tau\rangle_r\phi]{\Ge})\ dr > q,
\end{equation*}
the latter integral being equal to
$
\lambda(\{r \in [0, 1] \mid \Ge, s \models \langle\tau\rangle_r\phi\}).
$
Hence by~(\ref{lgth}) it follows that $\Ge, s \models \langle\tau\rangle_q\phi$. Thus we have found
\begin{equation*}
\Ge, s \models \langle{\posTest[p]};\tau\rangle_q\phi \Leftrightarrow 
\Ge, s \models p \wedge \langle\tau\rangle_q\phi.
\end{equation*}
Replacing in the above argumentation $V_p$ by $S \setminus V_p$, we see that $\Ge, s \models \langle{\negTest[p]};\tau\rangle_q\phi$ is equivalent to 
\begin{equation*}
D(s)(S\setminus V_p) = 1
\text{ and }
\int_0^1D(s)(\Gilt[\langle\tau\rangle_r\phi]{\Ge})\ dr > q.
\end{equation*}
Because we do not have negation in our logic, we obtain 
\begin{equation*}
\Ge, s \models \langle{\negTest[p]};\tau\rangle_q\phi \Leftrightarrow 
\Ge, s \not\models p \text{ and } \Ge, s \models\langle\tau\rangle_q\phi.
\end{equation*}

\section{Completion}
\label{sec:complete}
\def\cpl#1{\overline{#1}}
\def\rstr#1{\lfloor#1\rfloor}
We will deal now with the completion of an effectivity function, i.e., with its extension to the universal completion of the underlying state space. Universally complete measure spaces are closed under the Souslin operation by Proposition~\ref{closed-under-Souslin}. This in turn is required when we want to establish measurability of $\Gilt{}$ for an arbitrary formula $\phi$, see Proposition~\ref{prop-interp}. 

Completing the state space can be necessary, e.g., when working in a Polish space, because as a rule the measurable space generated from an uncountable  Polish spaces is not complete. This can be seen as follows. Let $S$ be an uncountable Polish space, then there exists an analytic set $A \subseteq S$ which is not a Borel set~\cite[Theorem XIII.11]{Kuratowski-Mostowski}. $A$ can be obtained through the Souslin operation  $\mathfrak{A}$ as
\begin{equation*}
A = \mathfrak{A}(\{F_v\mid v \in \wrd{\Nat}\})
\end{equation*}
(cp.~(\ref{Souslin})) with a family $\{F_v\mid v \in \wrd{\Nat}\}$  of closed sets. If $S$ would be complete, it would be closed under the Souslin operation by Proposition~\ref{closed-under-Souslin}, hence $A$ would be a Borel set, contrary to the assumption.

\subsection{Completing a Measurable Space}
\label{sec:the-compl}

Let $S$ be a measurable space, and define for $\mu\in\SubProb{S}$ and $A \subseteq S$ these well known set functions
\begin{itemize}
\item the \emph{inner measure} $\mu_*(A) := \sup\{\mu(B) \mid B \in \Borel{S}, B \subseteq A\}$,
\item the \emph{outer measure} $\mu^*(A) := \inf\{\mu(B) \mid B \in \Borel{S}, B \supseteq A\}$.
\end{itemize}
If $A \in \Borel{S}$ the inner measure $\mu_*(A)$ and the outer measure $\mu^*(A)$ coincide. 
The \emph{universal completion $\mathcal{U}$ of $S$} and the extension $\cpl\mu$ to $\mathcal{U}$ of $\mu\in\SubProb{S}$ is then defined through
\begin{align*}
\mathcal{U} & := \bigcap_{\nu\in\SubProb{S}} \{A \subseteq S \mid \nu_*(A) = \nu^*(A)\}\\
\cpl\mu(B) & := \mu^*(B)\text{ if } B \in \mathcal{U}.
\end{align*}
This space is usually denoted by $\cpl{S}$, so we put $\Borel{\cpl{S}} := \mathcal{U}$, it is plain that $\Borel{S}\subseteq\Borel{\cpl{S}}$. Thus the carrier set of the universal completion $\cpl{S}$ is the carrier set of the originally given space $S$, and $\Borel{S} \subseteq \Borel{\cpl{S}}$. Consequently, the identity $id: \cpl{S}\to S$ is $\Borel{\cpl{S}}$-$\Borel{S}$ measurable. Iterating the completion will not give any new results. The properties needed here are summarized:

\BeginProposition{completion}
Let  $S$ and $T$ be measurable spaces.
\begin{enumerate}
\item\label{completion-1} The universal completion $\cpl{S}$ is closed under the Souslin operation.
\item\label{completion-2} Given $\mu\in\SubProb{S}$, there exists a unique extension $\cpl\mu\in\SubProb{\cpl{S}}$. If $S$ is a metric space, then 
\begin{equation}
\label{outer}
\cpl\mu(B) = \inf\{\mu(G) \mid B \subseteq G, G \text{ open}\}
\end{equation}
holds for all $B\in\Borel{\cpl{S}}$. 
\item\label{completion-3}
If $f: S \to T$ is measurable, then $f: \cpl{S}\to\cpl{T}$ is measurable.
\end{enumerate}
\EndProposition

\BeginProof
\cite[Theorem 14.F]{Halmos},~\cite[Theorem 1.1.5, Theorem 7.1.7]{Bogachev}, see also~\cite[Section 7]{EED-PDL-TR} and~\cite[Proposition 4.3]{EED-PDL-TR}.  
\EndProof

The identity $id: \cpl{S}\to S$ is measurable, but we can say more

\BeginLemma{cpl-is-measur-1}
Let $S$ be a measurable space, then $\SubProb{id}: \SubProb{\cpl{S}}\to\SubProb{S}$ is bijective and measurable. 
\EndLemma

\BeginProof
1.
Given $\mu\in\SubProb{\cpl{S}}$, we have 
$
\SubProb{id}(\mu)(A) = \mu(A)
$
for $A \in \Borel{S}$, thus $\SubProb{id}(\mu)$ is the restriction of $\mu$ to the $\sigma$-algebra $\Borel{S}$.  Note that $\SubProb{\cpl{S}}$ is endowed with the $\sigma$-algebra which is generated by the sets
\begin{equation*}
\{\basS{\cpl{S}}{B}{> q} \mid B \in \Borel{\cpl{S}}\}.
\end{equation*}
Because 
$
\InvBild{\SubProb{id}}{\basS{S}{A}{> q}} = \basS{\cpl{S}}{A}{> q}
$
for $A\in\Borel{S}$, we infer measurability from $\cpl\mu$ being an extension to $\mu\in\SubProb{S}$.

2.
Assume $\SubProb{id}(\mu_1) =\SubProb{id}(\mu_2)$ for $\mu_1, \mu_2\in\SubProb{\cpl{S}}$, thus we have 
$
\mu_1(A) = \mu_2(A)
$
for all $A\in\Borel{S}$, hence $\mu_1 = \cpl{\SubProb{id}(\mu_1)} = \cpl{\SubProb{id}(\mu_2)} = \mu_2$ due to the uniqueness of the extension. 

3.
Given $\nu\in\SubProb{S}$, it is obvious that 
$
\nu = \SubProb{id}(\cpl\nu).
$
Thus $\SubProb{id}$ is onto.
\EndProof

\medskip

We know that $\SubProb{id}$ is bijective and measurable, but we cannot in general establish that its inverse is measurable as well. If we have a second countable metric space, however, we can say more. In this case we have a countable basis for the open sets, and from Proposition~\ref{completion}, \ref{completion-2}. we infer that can approximate $\cpl\mu(A)$ for arbitrary $A\in\Borel{\cpl{S}}$ by the values $\mu(G)$ for $G$ open. Investigating this closer, we find that we can represent the values of Borel sets through a suitable Souslin scheme of Borel sets in $\SubProb{S}$, which in turn permits the following observation.

\BeginLemma{cpl-is-measur-2}
If $S$ is a second countable metric space, then $\SubProb{id}^{-1}: \SubProb{S}\to\SubProb{\cpl{S}}$ is $\Borel{\cpl{\SubProb{S}}}$-$\Borel{\cpl{S}}$-measurable.
\EndLemma

\BeginProof
0.
The claim says that 
\begin{equation*}
\Bild{\SubProb{id}}{W} = \{\SubProb{id}(\mu) \mid \mu \in W\} = \{\rstr\mu \mid \mu \in W\}\in \Borel{\cpl{\SubProb{S}}},
\end{equation*}
whenever $W \in \bS{\cpl{S}}$. Here $\rstr\mu := \SubProb{id}(\mu)$ denotes the restriction of $\mu\in\SubProb{\cpl{S}}$ to $\Borel{S}$, thus $\cpl{\rstr\mu} = \mu$  for $\mu\in\SubProb{\cpl{S}}$ and $\rstr{\cpl\nu} = \nu$ for $\nu\in\SubProb{S}$. 

1.
Using~(\ref{outer}) and the fact that the open sets in $S$ are countable generated, we construct with exactly the same arguments as those in~\cite[Section 7.1]{EED-PDL-TR} for each
$
A\in\Borel{\cpl{S}}
$
a family
$
\{G_v\mid v \in \Nat^*\} \subseteq \Borel{\SubProb{S}}
$
such that 
\begin{equation*}
\Bild{\SubProb{id}}{\basS{\cpl{S}}{A}{\leq q}}
=
\{\nu\in\SubProb{S}\mid \nu^*(A) \leq q\}
= 
\mathfrak{A}(\{G_v\mid v \in \Nat^*\}).
\end{equation*}
Because $\cpl{\SubProb{S}}$ is universally complete, it is closed under the Souslin operation, which in turn implies that 
$
\Bild{\SubProb{id}}{\basS{\cpl{S}}{A}{\leq q}}\in\Borel{\cpl{\SubProb{S}}}.
$

2.
Because for all $A\in\Borel{\cpl{S}}$
\begin{equation*}
\basS{\cpl{S}}{A}{\leq q}\in\{W \in \bS{\cpl{S}} \mid \Bild{\SubProb{id}}{W}\in\Borel{\cpl{\SubProb{S}}}\},
\end{equation*}
and the latter set is a $\sigma$-algebra, we conclude that the assertion is true.
\EndProof

Call a measurable space $S$ \emph{separable} iff there exists a countable generator $\mathcal{C}$ for $\Borel{S}$ which separates points, so for any two distinct points there exists an element of $\mathcal{C}$ which contains  exactly one of them. It is well known that a measurable space $S$ is separable iff there exists a second countable metric topology $\tau$ on $S$ with $\Borel{S} = \sigma(\tau)$~\cite[Proposition 12.1]{Kechris}. Thus separable measurable spaces and second countable metric spaces are equivalent as measurable spaces. Consequently, Lemma~\ref{cpl-is-measur-2} holds also for separable spaces. The reason for formulating this statement for metric spaces is of course the existence of open sets and the regularity of measures.

\subsection{Completing a Model}
\label{sec:some-compl}

Let for the remainder of this section the state space $S$ be a separable space. We fix  a stochastic effectivity function $P$ on $S$ first, investigate its completion, and extend this result then to a game model.

The following definition formulates a property under which $P$ can be extended to a stochastic effectivity function $\cpl{P}$ on the completion $\cpl{S}$. 

\BeginDefinition{cplt-able}
The stochastic effectivity function $P$ is called \emph{completable} iff 
\begin{equation*}
\{\langle s, t\rangle \mid G_t\in P(s)\}\in \Borel{\cpl{S}\otimes[0, 1]},
\end{equation*}
whenever 
$
G \in \Borel{\cpl{\SubProb{S}}\otimes[0, 1]}.
$
\EndDefinition

Thus a completable effectivity function is guaranteed to handle the completion of the space of all subprobabilities gracefully by carrying t-measurability to the completion of the state space. 

As the name suggests, completable effectivity functions have an extension to the completion of the measurable space.

\BeginProposition{compl-exists}
Let $S$ be a separable space and $P$ a completable stochastic effectivity function on $S$. Then there exists a unique stochastic effectivity function $\cpl{P}$ in $\cpl{S}$ such that the identity $id: \cpl{P}\to P$ is a morphism.
\EndProposition

\BeginProof
1.
Define 
\begin{equation*}
\cpl{P}(s) := \{W \in \bS{S} \mid \rstr{W}\in P(s)\}
\end{equation*}
for $s\in S$, where $\rstr{W} := \{\rstr\mu\mid \mu\in W\}$ is the set of all restrictions of measures in $W \subseteq\SubProb{\cpl{S}}$. 
Then $\cpl{P}(s)$ is upper closed, and t-measurability remains to be established.

2.
We infer from Lemma~\ref{cpl-is-measur-2} that the map
\begin{equation*}
f:
\begin{cases}
\SubProb{S}\times[0, 1]&\to\SubProb{\cpl{S}}\times[0, 1]\\
\langle \mu,t \rangle&\mapsto \langle \cpl\mu,t \rangle
\end{cases}
\end{equation*}
is $\Borel{\cpl{\SubProb{S}}\otimes[0, 1]}-\Borel{\SubProb{\cpl{S}}\otimes[0, 1]}$-measurable. Thus if
$
H \in \Borel{\SubProb{\cpl{S}}\otimes[0, 1]}
$
is a test set, then 
\begin{equation*}
\widetilde{H} := \InvBild{f}{H} 
= \{\langle\rstr\mu, t\rangle\mid \langle\mu, t \in H\rangle\}\in \Borel{\cpl{\SubProb{S}}\otimes[0, 1]}.
\end{equation*}
Consequently,
\begin{equation*}
\{\langle s, t\rangle \mid H_t\in\cpl{P}(s)\}
=
\{\langle s, t\rangle \mid \widetilde{H}_t\in P(s)\}
\in \Borel{\cpl{S}\otimes[0, 1]},
\end{equation*}
because $P$ is completable. Consequently, $\cpl{P}$ is t-measurable.
\EndProof

\paragraph{Stochastic Relations.}
Let $K: S \Trans S$ be a stochastic relation, then we know from~\cite[Corollary 7.6]{EED-PDL-TR} that $K$ has a unique extension $\cpl{K}: \cpl{S}\Trans\cpl{S}$ with $\cpl{K}(s) = \cpl{K(s)}$. We will study briefly the relationship of $P_{\cpl{K}}$ and $\cpl{P_K}$, where $P_K$ is as above the stochastic effectivity function associated with $K$.

\BeginLemma{k-is-compl}
Let 
$
G \in \Borel{\cpl{\SubProb{S}}\otimes[0, 1]},
$
and put 
\begin{equation*}
\Omega(G) := \{\langle s, t\rangle\mid K(s)\in G_t\}
\end{equation*}
for the stochastic relation $K: S \Trans S$.
Then 
$
\Omega(G) \in \Borel{\cpl{S}\otimes[0, 1]}.
$
\EndLemma

\BeginProof
1.
Because
\begin{equation*}
\mathcal{X} := \{G \in \Borel{\cpl{\SubProb{S}}\otimes[0, 1]} \mid \Omega(G) \in \Borel{\cpl{S}\otimes[0, 1]}\}
\end{equation*}
is a $\sigma$-algebra (cp. (\ref{a-frege})), it is enough to show that 
$
\Omega(H\times I) \in \Borel{\cpl{S}\otimes[0, 1]},
$
when $H \in \Borel{\cpl{\SubProb{S}}}$ and $I \in\Borel{[0, 1]}$. This is so because these measurable rectangles generate the product $\sigma$-algebra. 

2.
Because $K: S \to \SubProb{S}$ is measurable, we infer from Proposition~\ref{completion}, part~\ref{completion-3}, that $K: \cpl{S}\to\cpl{\SubProb{S}}$ is measurable, consequently, 
$
\{s\in S \mid K(s)\in H\} \in\Borel{\cpl{S}}.
$
Thus
\begin{equation*}
\Omega(H\times I)
 = \{s \in S \mid K(s)\in H\} \times I
 = \InvBild{K}{H}\times I \in \Borel{\cpl{S}\otimes[0, 1]}.
\end{equation*}
Consequently, 
$
\mathcal{X} = \Borel{\cpl{\SubProb{S}}\otimes[0, 1]}.
$
\EndProof

This has as an immediate consequence
\BeginCorollary{pk-is-compl}
Let $P_K$ be the stochastic effectivity function associated with stochastic relation $K: S \Trans S$, then $P_K$ is completable.
\EndCorollary

\BeginProof
This follows from Lemma~\ref{k-is-compl} since
$
G \in P_K(s)
$
iff
$
K(s)\in G.
$
\EndProof

It might be noted that both Lemma~\ref{k-is-compl} and Corollary~\ref{pk-is-compl} do not depend on topological assumptions; they hold in general measurable spaces as well. 

It is now easy to compare the completion of a stochastic effectivity function associated with a stochastic relation to the stochastic effectivity function associated with the completion of a stochastic relation.

\BeginProposition{compare-eff-fncts}
Let $K: S \Trans S$ be a stochastic relation with extension $\cpl{K}:\cpl{S}\Trans\cpl{S}$. Then
\begin{equation*}
P_{\cpl{K}} = \cpl{P_K}.
\end{equation*}
\EndProposition

\BeginProof
Because $P_K$ is completable by Corollary~\ref{pk-is-compl}, $\cpl{P_K}$ is a stochastic effectivity function on $\cpl{S}$. Then we have 
\begin{equation*}
W \in P_{\cpl{K}}(s)
\Leftrightarrow
\cpl{K}(s) \in W
\Leftrightarrow
K(s) \in \rstr{W}
\Leftrightarrow
W \in \cpl{P_K}(s).
\end{equation*}
\EndProof

\paragraph{Completing a Game Model.}
With Proposition~\ref{compl-exists} we are in a position to complete a game model over a separable space, provided the effectivity functions for the primitive games are well behaved.

\BeginProposition{compl-model}
Let $\Ge = (S, (P_\gamma)_{\gamma\in\Gamma}, (V_p)_{p\in \AE})$ be a game model over a separable space $S$ such that $P_\gamma$ is completable for all $\gamma\in\Gamma$. Then there exists a unique game model $\cpl{\Ge}$ over the completion $\cpl{S}$ such that $id: \cpl{\Ge}\to\Ge$ is a model morphism.
\EndProposition

\BeginProof
Define 
$
\cpl{\Ge} := (\cpl{S}, (\cpl{P_\gamma})_{\gamma\in\Gamma}, (V_p)_{p\in \AE}),
$
then $\cpl{\Ge}$ is a game model such that $id: \cpl{P_\gamma}\to P_\gamma$ is a morphism. Because 
$
\InvBild{id}{V_p} = V_p \in \Borel{\cpl{S}},
$
we infer that $id: \cpl{\Ge}\to\Ge$ is a model morphism. Uniqueness of $\cpl{\Ge}$ is trivial.
\EndProof

If $\Ke$ is a stochastic Kripke model, then $\Ke$ can be extended in an obvious way to a unique stochastic Kripke model $\cpl{\Ke}$ over $\cpl{S}$ such that the identity is a morphism $\cpl\Ke\to\Ke$. Because the effectivity function associated with a stochastic relation is completable, we obtain from Proposition~\ref{compl-model} through Proposition~\ref{compare-eff-fncts}:

\BeginProposition{compl-vs-compl}
Let $\Ke$ be a stochastic Kripke model over a separable space $S$, and denote the associated game model by $\Ge_\Ke$. Then 
$
\cpl{\Ge_\Ke}
$
is identical to 
$
\Ge_{\cpl\Ke}.
$
\QED
\EndProposition

Thus, if we are given a stochastic Kripke model over a separable space $S$, we can interpret game logic over the completion of this state space, obtaining universally measurable validity sets for all formulas, in particular those which involve angelic or demonic iteration. If, however, we are given a general game model over $S$, we have to take care that its effectivity functions behave decently to completing the state space. If this is the case, we are complete the model and obtain universally measurable validity sets as well.  
 
\section{Conclusion and Further Work}
\label{sec:concl}
An interpretation of game logic through models based on stochastic effectivity functions is proposed. These functions are investigated, and their relationship with stochastic relations is characterized completely. Morphisms and congruences for effectivity functions are defined, they are used for an investigation of logical and behavioral equivalence of game models. It is finally shown that a model can be completed by constructing the universal completion of the underlying measurable space.

\paragraph{Further Work.} The effectivity functions defined here serve as generalizations of stochastic relations, on which the model of Markov transition systems is based. The interplay between them is characterized completely in general measurable spaces. Because they have a fairly non-deterministic character ---~their range consisting of portfolios, i.e., of measurable sets of measures~--- they may be used as a model for stochastic non-determinism as well; this has been proposed in\cite{Terraf-Bisim-MSCS}. The nondeterministic Markov decision processes discussed in tat paper are based on measurable maps, but with a much weaker concept of measurability, suggesting further works for modelling non-determinism with these functions. This becomes probably even more interesting when topological assumptions are made, e.g., when the state space is a Polish space, and the portfolios are closed or even compact sets of distributions. Then measurability becomes an issue again, because t-measurability as defined here has to compete with Borel measurability; conditions under which measurable selectors exist might be of interest as well. Such selectors could be of interest as models for randomized policies along the lines of dynamic optimization~\cite{Hinderer}. 

Expressivity of models did not touch bisimilarity, which of course is a topic of interest of its own. Parikh and Pauly~\cite{Pauly-Parikh} extend the notion of bisimilarity to their models; this is patterned after bisimilarity of Kripke models. It is shown in~\cite{EED-Game-Coalg} that this can be treated coalgebraically by translating the scenario of effectivity functions into a coalgebraic one. Two models are shown to be bisimilar in the sense of Parikh and Pauly iff there exists a model based on a relation between the state spaces such that the corresponding projections are morphisms. A corresponding characterization for the stochastic situation would be interesting. It would be even more interesting to observe the interplay with logical and behavioral equivalence. Here certainly additional assumptions on the structure of the state spaces involved are necessary, and most likely additional tools from set theory are required.  

\bibliographystyle{plain}

\end{document}